\documentclass[journal]{IEEEtran}

\usepackage[table,xcdraw]{xcolor}

\usepackage{amsmath,amsfonts}
\usepackage{algorithmic}
\usepackage{algorithm}
\usepackage{array}
\usepackage[caption=false,font=normalsize,labelfont=sf,textfont=sf]{subfig}
\usepackage{textcomp}
\usepackage{stfloats}
\usepackage{url}
\usepackage{verbatim}
\usepackage{graphicx}
\usepackage{cite}

\usepackage{hhline}
\usepackage{adjustbox}
\usepackage{float}
\usepackage{booktabs}
\usepackage{multirow}
\usepackage{tabu}
\usepackage{lscape}
\usepackage[normalem]{ulem}

\usepackage{lettrine}
\usepackage{verbatim}
\usepackage[hidelinks]{hyperref}
\usepackage{url}
\usepackage{lscape}
\usepackage{comment, soul}
\usepackage{bm}

\usepackage{diagbox}
\usepackage{slashbox}
\usepackage{makecell}
  
\newcolumntype{C}{>{\centering\arraybackslash}X}
\usepackage{pifont}
\definecolor{redtable}{RGB}{255, 153, 153}
\definecolor{greentable}{RGB}{128, 255, 128}
\usepackage{bm}

\usepackage{cite}

\usepackage{hyperref}
\hypersetup{
    colorlinks=true,
    linkcolor=blue,
    filecolor=magenta,      
    urlcolor=cyan,
}

\newcommand\notsotiny{\@setfontsize\notsotiny{6}{7}}
\newcolumntype{x}[1]{>{\centering\arraybackslash\hspace{0pt}}p{#1}}

\def\BibTeX{{\rm B\kern-.05em{\sc i\kern-.025em b}\kern-.08em
    T\kern-.1667em\lower.7ex\hbox{E}\kern-.125emX}}

\usepackage{array, makecell}
\usepackage{adjustbox}
\makeatletter
\def\hlinewd#1{%
\noalign{\ifnum0=`}\fi\hrule \@height #1 %
\futurelet\reserved@a\@xhline}
\makeatother

\newcolumntype{L}[1]{>{\raggedright\let\newline\\\arraybackslash\hspace{0pt}}m{#1}}
\newcolumntype{C}[1]{>{\centering\let\newline\\\arraybackslash\hspace{0pt}}m{#1}}
\newcolumntype{R}[1]{>{\raggedleft\let\newline\\\arraybackslash\hspace{0pt}}m{#1}}

\usepackage{caption}
\newcommand{\cmark}{\ding{51}}%
\newcommand{\xmark}{\ding{55}}
\usepackage{acronym}
\usepackage[normalem]{ulem}
\useunder{\uline}{\ul}{}
\newacro{iqm}[IQM]{Image Quality Metric}
\newacro{st-rred}[ST-RRED]{Spatio-Temporal Reduced Reference Entropic Differencing}
\newacro{tlvqm}[TLVQM]{Two-Level Approach for No-Reference Consumer Video Quality Assessment}
\newacro{iqa}[IQA]{Image Quality Assessment}
\newacro{cidiq}[CIDIQ]{colourlab image database image quality}
\newacro{vdid}[VDID]{viewing distance-changed image database}
\newacro{live}[LIVE]{laboratory for image and video engineering}
\newacro{mlive}[M-LIVE]{multi-distance laboratory for image and video engineering}
\newacro{gru}[GRU]{Gated Recurrent Unit}
\newacro{rapique}[RAPIQUE]{Rapid and Accurate Video Quality Prediction Evaluator}
\newacro{lcf}[LCF]{Low Complexity Features}
\newacro{hcf}[HCF]{High Complexity Features}
\newacro{videval}[VIDEVAL]{VIDeo quality EVALuator}
\newacro{vqm}[VQM]{Video Quality Model}
\newacro{itu}[ITU]{International Telecommunication Union}
\newacro{frcnn}[Fast-RCNN]{Fast Region Based Convolutional Neural Networks}

\newacro{fr}[FR-VQA]{Full Reference Video Quality Assessment}
\newacro{nr}[NR-VQA]{No Reference Video Quality Assessment}
\newacro{rr}[RR-VQA]{Reduced Reference Video Quality Assessment}
\newacro{cnn}[CNN]{Convolutional Neural Network}
\newacro{rnn}[RNN]{Recurrent Neural Network}
\newacro{nn}[NN]{Neural Network}
\newacro{vmaf}[VMAF]{Video Multi-method Assessment Fusion}
\newacro{mtl}[MTL]{Multi-task Learning}
\newacro{mse}[MSE]{Mean Squared Error}
\newacro{mos}[MOS]{Mean Opinion Score}
\newacro{psnr}[PSNR]{Peak Signal to Noise Ratio}
\newacro{hvs}[HVS]{Human Visual System}
\newacro{ssim}[SSIM]{Structural SIMilarity}
\newacro{msssim}[MS-SSIM]{multi-scale structural similarity index}
\newacro{mlp}[MLP]{Multi-layer Perceptron}
\newacro{lgn}[LGN]{Lateral Geniculate Nucleus}
\newacro{ifc}[IFC]{image fidelity criterion}
\newacro{vif}[VIF]{Visual Information Fidelity}
\newacro{nss}[NSS]{Natural Scene Statistics}
\newacro{gsm}[GSM]{Gaussian Scale Mixture}
\newacro{diivine}[DIIVINE]{Distortion Identification-based Image Verity and Integrity evaluation}
\newacro{niqe}[NIQE]{Natural Image Quality Elevator}
\newacro{bliinds}[BLIINDS]{Blind Image Integrity Notator using DCT Statistics}
\newacro{bliinds2}[BLIINDS-\RomanNumeralCaps 2]{blind Image integrity notator using DCT statistics}
\newacro{dct}[DCT]{Discrete Cosine Transform}
\newacro{ann}[ANN]{Artificial Neural Network}
\newacro{grnn}[GRNN]{General Regression Neural Network}
\newacro{svr}[SVR]{Support Vector Regression}
\newacro{fc}[FC]{Fully Connected}
\newacro{deepbiq}[DeepBIQ]{Deep Blind Image Quality}
\newacro{lts}[LTS]{Long Term Support}
\newacro{spp}[SPP]{Spatial Pyramid Pooling}
\newacro{roip}[RoIPool]{Region of Interest Pooling}
\newacro{pri}[PRI]{Pseudo-Reference Image}
\newacro{dct}[DCT]{Discrete Cosine Transform}
\newacro{sota}[SOTA]{State-Of-The-Art}
\newacro{dwt}[DWT]{Discrete Wavelet Transform}
\newacro{bnbv}[BNBV]{blurriness noisiness blockiness viewing distance}
\newacro{svm}[SVM]{Support Vector Machine}
\newacro{mvg}[MVG]{Multivariate Gaussian}
\newacro{rf}[RF]{Random Forest}
\newacro{dcnn}[DCNN]{dense convolutional neural network}
\newacro{db}[DB]{dense block}
\newacro{tb}[TB]{transition block}
\newacro{kadid}[KADID-10K]{konstanz artificially distorted image quality database}
\newacro{fsim}[FSIM]{feature similarity index}
\newacro{vsi}[VSI]{visual saliency induced}
\newacro{gmsd}[GMSD]{gradient magnitude similarity deviation}
\newacro{brisque}[BRISQUE]{Blind/Referenceless Image Spatial Quality Evaluator}

\newacro{plcc}[PLCC]{Pearson Linear Correlation Coefficient}
\newacro{srocc}[SROCC]{Spearman Rank Order Correlation Coefficient}
\newacro{krcc}[KRCC]{Kendall Rank Correlation Coefficient}
\newacro{rmse}[RMSE]{Root Mean Squared Error}

\newacro{dmos}[DMOS]{Differential Mean Opinion Score}
\newacro{gpu}[GPU]{Graphics Processing Unit}
\newacro{ugc}[UGC]{User-Generated Content}
\newacro{pgc}[PGC]{Professionally Generated Content}
\newacro{lstm}[LSTM]{Long Short Term Memory}
\newacro{bi-lstm}[Bi-LSTM]{Bi-directional Long Short Term Memory}
\newacro{bvqa}[BVQA]{Blind Video Quality Assessment}
\newacro{biqa}[BIQA]{Blind Image Quality Assessment}
\newacro{vqa}[VQA]{Video Quality Assessment}
\newacro{gap}[GAP]{Global Average Pooling}
\newacro{insla}[INSLA]{Iterative Nested Least Squares Algorithm}
\newacro{viideo}[VIIDEO]{Video Intrinsic Integrity and Distortion Evaluation Oracle}

\begin{document}

\title{2BiVQA: Double Bi-LSTM based Video Quality Assessment of UGC Videos}

\author{Ahmed Telili, Sid Ahmed Fezza, Wassim Hamidouche,~\IEEEmembership{Member,~IEEE} and Hanene F. Z. Brachemi Meftah
\thanks{A. Telili and W. Hamidouche  are with Univ. Rennes, INSA Rennes, CNRS, IETR - UMR 6164, Rennes, France (e-mail: \href{mailto:atelili@insa-rennes.fr}{atelili@insa-rennes.fr} and \href{mailto:whamidouche@insa-rennes.fr}{whamidouche@insa-rennes.fr} ).}
\thanks{SA. Fezza and H. F. Z. Brachemi Meftah are with National Higher School of Telecommunications and ICT, Oran, Algeria (e-mail: \href{mailto:sfezza@ensttic.dz}{sfezza@ensttic.dz}).}}

\markboth{ }%
{Shell \MakeLowercase{\textit{et al.}}: A Sample Article Using IEEEtran.cls for IEEE Journals}


\maketitle

\begin{abstract}
  Recently, with the growing popularity of mobile devices as well as video sharing platforms (e.g., YouTube, Facebook, TikTok, and  Twitch), \ac{ugc} videos have become increasingly common and now account for a large portion of multimedia traffic on the internet. Unlike professionally generated videos produced by filmmakers and videographers, typically, \ac{ugc} videos contain  multiple authentic distortions, generally introduced during  capture and processing by naive users. Quality prediction of \ac{ugc} videos is of paramount importance to optimize and monitor  their processing in hosting platforms, such as their coding, transcoding, and streaming. However, blind quality prediction of \ac{ugc} is quite challenging because the degradations of \ac{ugc} videos are unknown and very diverse, in addition to the unavailability of pristine reference. Therefore, in this paper, we propose an accurate and efficient \ac{bvqa} model for \ac{ugc} videos, which we name 2BiVQA for double Bi-LSTM Video Quality Assessment. 2BiVQA metric consists of three main blocks, including a pre-trained \ac{cnn} to extract discriminative  features from image patches, which are then fed into two \acp{rnn} for spatial and temporal pooling. Specifically, we use two \ac{bi-lstm} networks, the first is used to capture short-range dependencies between image patches, while the second allows capturing long-range dependencies between frames to account for the temporal memory effect. Experimental results on recent large-scale \ac{ugc} \acs{vqa} datasets show that 2BiVQA achieves high performance at lower computational cost than most state-of-the-art \acs{vqa} models. The source code of our 2BiVQA metric is made publicly available at: \href{https://github.com/atelili/2BiVQA}{https://github.com/atelili/2BiVQA.}
\end{abstract}

%

\acresetall

\section{Introduction}
Currently, video represents the majority of Internet traffic. According to Cisco's recent report~\cite{cisco2020cisco}, it now accounts for around $82\%$ of global Internet traffic. Some of this traffic is generated by streaming video providers like Netflix, Amazon Prime Video, etc. Usually, the content they provide has been created by experts using professional capture devices and in a controlled environment, known as \ac{pgc}. \ac{pgc} videos are pristine high-quality videos that reach a certain level of perfection, making them suitable candidates as references in \ac{vqa} process. On the other hand, \ac{ugc} accounts for a significant portion of video traffic, which is collected and shared over social media and other video-sharing platforms, such as Facebook, Youtube, TikTok and Twitch. This content is typically captured by nonprofessional users using their own capture devices (e.g., smartphones) and under different shooting conditions. Unlike \ac{pgc} videos, \ac{ugc} videos may suffer from multiple authentic distortions that can be introduced during acquisition. Moreover, compression and transmission distortions are still introduced before uploading to the hosting platform. \ac{ugc} distortions are unpredictable, more diverse, intermixed, and the unavailability of a pristine reference makes the prediction of \ac{ugc} video quality very challenging. Thus, there is a great need to accurately assess the quality of \ac{ugc} videos in order to optimize and monitor their processing in hosting platforms, such as their coding, transcoding and streaming.

For \ac{vqa}, the most reliable technique is to perform a subjective quality evaluation. In subjective tests, a panel of human viewers is asked to rate the quality of stimuli displayed and assessed under a particular protocol and viewing conditions~\cite{bt2012500}. However, subjective tests are time-consuming, costly, and they cannot be used in real-time applications. As an alternative, objective quality measures have been developed to automatically predict the quality of videos. Depending on the required amount of the reference information, objective \ac{vqa} metrics can be divided into three categories: \ac{fr}, \ac{rr} and \ac{nr}. \Ac{fr} quality metrics require the presence of the entire pristine video frames to compare against in order to compute the quality score. However, adopting such a strategy for \ac{ugc} videos is not consistent, since the videos uploaded to the hosting platform have already undergone distortions due to acquisition and compression, making them not suitable as reference videos. Thus, no reference or blind \ac{vqa} metrics remain the obvious solution that solves the \ac{ugc}-\ac{vqa} issue. Although most recent \ac{biqa}/\ac{bvqa} methods achieve good performance on synthetic distortion datasets~\cite{seshadrinathan2010study}, their performance on \ac{ugc} videos remains far from satisfactory~\cite{tu2021ugc,hosu2017konstanz,sinno2018large,wang2019youtube,xu2021perceptual}, and predicting the quality of \ac{ugc} videos is still challenging and unsolved problem. 


Recently, with the massive growth in social media, attention has moved more towards building an accurate and efficient \ac{bvqa} model suitable for \ac{ugc} content, which allows achieving more intelligent analysis and processing in various applications~\cite{zhang2015aesthetics}. Hence, in recent years, researchers have deployed considerable efforts into the development of in-the-wild \ac{ugc} datasets such as KoNViD-1k~\cite{hosu2017konstanz}, LIVE-VQC~\cite{sinno2018large}, and YouTube-UGC~\cite{wang2019youtube}, to cite a few examples. These \ac{ugc} datasets differ significantly from synthetic distortion datasets by a varied type and a wide range of distortions, but also by the fact that the distortion is not uniformly distributed over the spatial and temporal domains, resulting in fluctuating video quality.

The existing metrics do not consider or consider insufficiently this last aspect. They do not take into account how non-uniform distortions affect the overall frame quality score and how adjacent frames, from past and future, impact the perceived quality of the current frame. Typically, existing metrics use the mean as a pooling strategy, which is not a good representation of the spatial-temporal quality distributions. In this regard, we use a data-driven deep-learning approach in the proposed metric to enhance the \ac{vqa}.

It is obviously desirable to have accurate video quality metrics for the \ac{ugc} videos. Thus, in this work, we propose an efficient model for \ac{ugc}-\ac{vqa}, termed 2BiVQA for double Bi-LSTM Video Quality Assessment. The main contributions of this paper can be summarized
as follows:
\begin{itemize}
  \item We propose an accurate and efficient \ac{bvqa} metric for \ac{ugc} that performs the quality assessment in line with the \ac{hvs}. The components of 2BiVQA include a \ac{cnn} for spatial feature extraction and two \acp{rnn} for capturing spatial-temporal dependencies. We show that pre-training the features extraction module on an in-the-wild \ac{iqa} dataset significantly improves the performance of 2BiVQA.
  \item We leverage two \acp{rnn}, namely \ac{bi-lstm} networks, for both spatial and temporal pooling, which allows our model to take into account the characteristics of \ac{ugc} videos as well as the \ac{hvs} behavior.
  \item We conduct experiments on three  widely-used \ac{ugc}-\ac{vqa} datasets to demonstrate the effectiveness of 2BiVQA. Experimental results show that the proposed 2BiVQA metric achieves competitive performance with \ac{sota} methods and provides the best generalization capability, even at a low computational cost.

\end{itemize}

The rest of this paper is organized as follows. Section \ref{related work} reviews related work, then Section \ref{proposed} presents the proposed 2BiVQA model. The performance of our model is assessed and analyzed in Section \ref{exp}. Finally, Section \ref{cncls} concludes the paper.

\section{Related work}
\label{related work}
Given the unavailability of pristine sources, \ac{fr} metrics cannot well predict the perceptual quality of \ac{ugc} videos. Thus, in this section, we focus on \ac{bvqa} methods, as these methods are the most suitable for providing  \ac{ugc} video quality estimation. \ac{bvqa} methods can be grouped into two categories, whether their relevant features are extracted from the input video based on conventional handcrafted techniques or deep learning-based models.

\subsection{Handcrafted feature-based  BVQA models}
The earliest \ac{biqa}/\ac{bvqa} methods were mostly distortion-specific QA algorithms, which address a specific type of distortion, such as blur~\cite{marziliano2002no, wang2008blind}, blockiness~\cite{wang2000blind, min2017unified}, ringing~\cite{feng2006measurement}, banding~\cite{wang2016perceptual, tu2020bband} and noise~\cite{amer2005fast, norkin2018film} or targeted multiple types of distortion~\cite{gu2014hybrid,lu2015no, min2018blind}. 
Later on, the most successful handcrafted features-based \ac{bvqa} models mainly rely on learning approaches~\cite{zhai2020perceptual,min2021screen}, using a set of relevant perceptual features combined with a regression model to predict quality scores~\cite{moorthy2011blind, kundu2017no, ghadiyaram2017perceptual, pei2015image}.

The most popular \ac{biqa}/\ac{bvqa} algorithms are based on \ac{nss}~\cite{ruderman1994statistics}, extracted from either spatial domain or transform domain. \ac{nss} refer to the hypothesis that the natural scenes form a minor subspace within the space of all conceivable signals. These \ac{nss} are altered by the presence of distortions, so they were widely used to blindly measure the quality of images/videos. Successful models relying on \ac{nss} are derived from the spatial domain  (NIQE~\cite{mittal2012no}, BRISQUE~\cite{mittal2012no}), \ac{dct} (BLIINDS~\cite{saad2010dct} and  BLIINDS-II~\cite{saad2012blind}) and \ac{dwt} (BIQI~\cite{moorthy2010two}, DIIVINE~\cite{moorthy2011blind}, C-DIIVINE~\cite{zhang2014c}). These metrics have been expanded to  the \ac{vqa} task using the space-time natural video statistics models~\cite{li2016spatiotemporal, sinno2019spatio,saad2014blind}.
For instance, in \cite{liu2020blind}, an \ac{nss}-based method was proposed for \ac{biqa}, which consists in extracting \ac{nss} features from the local binary pattern (LBP) map and the mean subtracted and contrast normalized (MSCN) coefficients of the image. Thus, the extracted \ac{nss} features along with other features related to perceptual characteristics constitute the quality-aware features for quality estimation. 
Other extensions of \ac{nss} have been proposed including in log-derivative and log-Gabor spaces  (DESIQUE~\cite{zhang2013no}), the joint statistics of the gradient magnitude and Laplacian of Gaussian responses in the spatial domain (GM-LOG~\cite{xue2014blind}) and the gradient domain of LAB color transforms (HIGRADE~\cite{kundu2017no}). HIGRADE is based on both gradient orientation information extracted from the gradient structure tensor and gradient magnitude. Once these features are extracted, a mapping is performed from the feature space to the \ac{mos} scores using the \ac{svr}. In~\cite{ghadiyaram2017perceptual}, the authors proposed a bag of feature-maps approach. In which several feature maps are derived from multiple color spaces and transform domains, then scene statistics from each of these maps are extracted. The obtained results demonstrated the relevance of the extracted features for the quality prediction of images corrupted by complex mixtures of authentic distortions. Motivated by the success of unsupervised feature learning for \ac{biqa} in CORNIA~\cite{ye2012unsupervised}, the authors proposed to extend it to video signal V-CORNIA~\cite{xu2014no}. In V-CORNIA, frame-level features are extracted by unsupervised feature learning, and a \ac{svr} is subsequently used to learn a mapping from feature space to frame-level quality scores. Finally, an overall video quality score is derived via a hysteresis temporal pooling. Min \textit{et al.} \cite{min2017blind,min2018blind} introduced a new concept called \ac{pri} and developed a PRI-based BIQA framework (BPRI). Unlike traditional reference image, which is assumed to have a perfect quality, PRI is generated from the same distorted image and intentionally subjected to the highest distortion. Notably, this framework employs PRI to predict image quality by formulating distortion-specific metrics that evaluate diverse types of distortion, including blockiness, sharpness and noisiness. These metrics evaluate the structural similarity between a distorted image and its corresponding PRI. Thus, a highly distorted image will have a higher degree of similarity with the corresponding PRI. The concept of PRI has been extended to Multiple PRIs (MPRI) in~\cite{min2018blind}, which are obtained by further degrading the distorted image in several ways and to certain degrees, and then comparing the similarities between the distorted image and its MPRIs.

To leverage these rich \ac{iqa} metrics for \ac{vqa} context, a straightforward approach is to compute the quality score of each frame and then pool them into an overall video quality score. The most adopted temporal pooling method is the average, however, this approach does not take into account the temporal change and quality fluctuation. This is why more advanced temporal pooling strategies have been proposed
~\cite{seshadrinathan2011temporal,xu2014no,tu2020comparative}.

However, performing \ac{vqa} cannot be based solely on spatial information, i.e., based only on \ac{iqa} metrics, since temporal information such as motion plays a crucial role in the perception of quality/distortion in the video and must be taken into account. Therefore, unlike simply extending \ac{iqa} methods to assess video quality using a pooling strategy, other methods have attempted to include temporal information directly in their models. For instance, a completely blind metric called \ac{viideo} was proposed in~\cite{mittal2015completely}. VIIDEO is based on a set of perceptually relevant temporal video statistic models of video frame difference signals. Inter-subband correlations over local and global time spans were used to quantify the degree of distortion in the video and thus predict the quality score. Manasa and Channappayya~\cite{manasa2016optical} proposed estimating perceptual quality by estimating statistical irregularities in optical flow using features at the patch and frame levels. V-BLIINDS has been proposed in~\cite{saad2014blind}, which includes a spatiotemporal \ac{nss} model of DCT coefficient statistics, as well as a motion model that quantifies motion coherency in the video. Li \textit{et al.}~\cite{li2016spatiotemporal} proposed a \ac{bvqa} based on the spatiotemporal statistics of videos in the 3D-DCT domain, which allows describing the spatial and temporal regularities of local space-time regions simultaneously. 
\Ac{tlvqm}~\cite{korhonen2019two} is another handcrafted features-based \ac{bvqa} method relying on a two-level approach for features extraction. First, \ac{lcf} are calculated at a rate of one frame per second over the entire video sequence, then the \acs{lcf} are utilized to select a set of representative subset of frames for calculating \ac{hcf}. Finally, both low and high complexity features are aggregated as a single feature vector representing the whole video sequence by using  \ac{svr} as a regression model. A more recent fusion-based \ac{bvqa} model is \ac{videval}~\cite{tu2021ugc}, which is based on features selection among top-performing \ac{biqa}/\ac{bvqa} models such as BRISQUE, HIGRADE, TLVQM, etc. To select the most relevant features, \ac{rf} is used to  remove the less significant features. Finally, a \ac{svm} with a linear kernel is trained to regress the final features vector into a quality score. Kancharla \textit{et al.}~\cite{9633248} proposed a \ac{bvqa} method, which includes a bandpass filter model of the visual system to evaluate the temporal quality and a weighted NIQE module to estimate the frame-level spatial quality. Finally, the global video quality score is computed by the average of the spatial quality and the temporal quality. All previous methods tried to predict the average quality perceived by end users, known as \ac{mos}. Differently, in~\cite{tiotsop2022mimicking}, the authors proposed to model how a single observer perceives the media quality using a neural network instead of predicting the \ac{mos}. The training of a neural network relies on the observer's ratings collected from subjective experiments to mimic his quality judgment, which implicitly accounts for his individual characteristics such as user expectations and personality that have an impact on quality of experience~\cite{zhu2018measuring}. 
\subsection{Deep learning-based BVQA models}
In recent years, deep \acp{cnn} have shown outstanding  performance in a wide range of computer vision tasks such as image classification~\cite{simonyan2014very,he2016deep}, object detection~\cite{bochkovskiy2020yolov4, girshick2015fast} and image segmentation~\cite{ronneberger2015u, liu2015parsenet}. Recently, with the release of several larger \ac{iqa}/\ac{vqa} datasets~\cite{ghadiyaram2015massive, wang2019youtube,gotz2021konvid, min2017unified}, deep \acp{cnn} have been extensively explored to solve image/video quality assessment problem. Yet, due to the lack of large-scale \ac{iqa}/\ac{vqa} datasets, it is quite challenging to train a deep \ac{cnn} from scratch to reach a competitive performance. To overcome the limitation of small data size, two solutions have been used in the literature: 1) performing a patch-wise training to increase data samples~\cite{kang2014convolutional, kim2017deep}, or 2) leveraging pre-trained deep \acp{cnn} on large datasets like ImageNet ~\cite{deng2009imagenet}, then performing fine-tuning on target \ac{iqa}/\ac{vqa} datasets.

\begin{figure*}[!t]
  \centering
  \centerline{\includegraphics[width=1\linewidth]{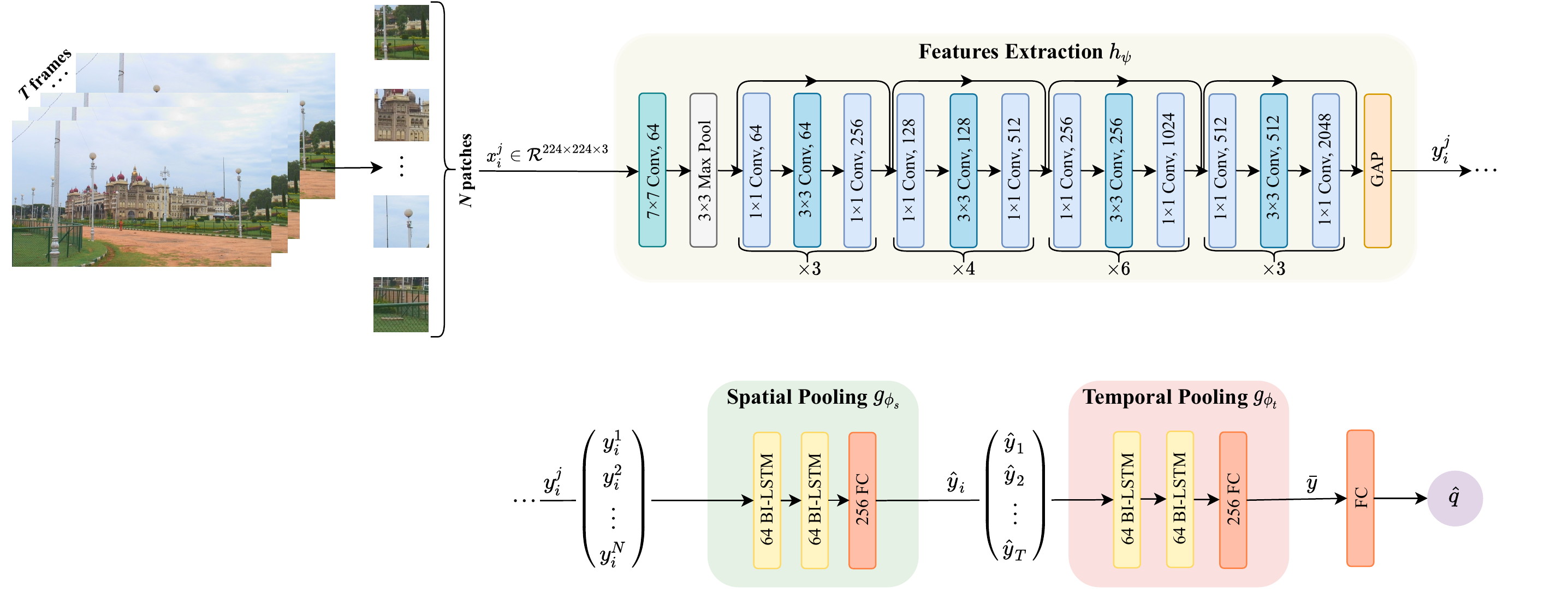}}

\caption{The overall framework of the proposed 2BiVQA metric. The features extraction module is used to extract spatial features $\bm{y}_i^j$ from patches $\bm{x}_i^j$. The spatial and temporal pooling modules are used to aggregate features into a final vector $\bar{\bm{y}}$ while accounting for \ac{hvs} behavior. Finally, the regression module uses the final vector $\bar{\bm{y}}$ to predict the quality score $\hat{q}$.}
\label{overall} \vspace{-4mm}
\end{figure*}

The first adoption of a \ac{cnn} model to the problem of \ac{iqa} was made by Kang \textit{et al.} in~\cite{kang2014convolutional}, where a \ac{cnn} was used to blindly predict the image quality score. They combined feature learning and regression in end-to-end optimization without using handcrafted features. Following this work, considerable deep learning-based \ac{biqa} methods have been proposed~\cite{yang2019survey, sun2023blind}, which achieved quite good performance. For video, on the other hand, very few methods based on deep learning have been dedicated to \ac{bvqa}.


For instance, Ahn  \textit{et al.}~\cite{ahn2018deep} proposed a \ac{bvqa} metric based on a deep \ac{cnn} model named DeepBVQA,  which includes various spatial and temporal cues. In DeepBVQA, a pre-trained \ac{cnn} model for \ac{iqa} is used to extract spatial features from each frame, and temporal sharpness variation is exploited  to extract temporal features. Then, these spatial and temporal features are combined into a video feature to be regressed to a final quality score. Another deep learning-based \ac{vqa} model was proposed in~\cite{you2019deep}, which consists of a 3D-\ac{cnn} to extract spatio-temporal features followed by a \ac{lstm} to predict the perceived video quality. A multi-task \ac{cnn} framework, named V-MEON, was proposed in~\cite{liu2018end} that predicts both the quality score and codec type of a video. V-MEON is based on 3D-\ac{cnn} network to extract spatio-temporal
features from a video, followed by the codec classifier and the quality predictor that are jointly optimized. VSFA~\cite{li2019quality} model also uses a \ac{cnn}, pre-trained on image classification tasks, as a features extraction module, and then it uses \ac{gru} and a subjectively-inspired temporal pooling layer to output the video quality score. Next, an improved version of VSFA, named MDVSFA, was proposed in~\cite{li2021unified}. MDVSFA uses a mixed dataset training strategy for training a single \ac{vqa} model with multiple datasets. Yi \textit{et al.}~\cite{yi2021attention} proposed a modified VGG-16 network with non-local layers to learn the global relationship of spatial features, which can be regarded as a kind of attention mechanism. In addition, they combined \ac{gru} and a temporal pooling layer to model the temporal-memory effects.

More recently~\cite{zhang2023md}, Zhang \textit{et al.} introduced a Multi-Dimensional VQA (MD-VQA) metric aimed at assessing the visual quality of compressed \ac{ugc} live videos. This method evaluates the quality in terms of semantic, distortion, and motion aspects. 
Tu \textit{et al.}~\cite{tu2021rapique}  proposed a hybrid method, named \ac{rapique}, which uses both handcrafted and deep \ac{cnn}-based high-level features. RAPIQUE is based on two modules, a \ac{nss} features extractor module, which extracts both spatial and temporal features, and a deep \ac{cnn} features extractor (ResNet-50) which extracts deep high-level features. Finally, a regressor model is used to map the extracted features to a quality score. Another deep learning-based \ac{vqa} model was proposed in \cite{sun2022deep}, including a 2D-\ac{cnn} to extract quality-aware spatial feature representation from raw pixels of the video frames, as well as a 3D-\ac{cnn} dedicated to the extraction of motion features, followed by a \ac{mlp} regression module to map these features into chunk-level quality scores, and finally, temporal average pooling is used to derive the video-level quality score. In~\cite{sun2021deep}, the authors proposed to hierarchically add the feature maps from intermediate layers into the final feature maps and calculate their global mean and standard deviation as the feature representation. Thus, covering the full range of visual features from low-level to high-level. Subsequently,  \ac{fc} and temporal pooling are used for the quality regression. In the same way, Shen \textit{et al.}~\cite{shen2022end} proposed a \ac{bvqa} method with spatio-temporal feature fusion and hierarchical information integration. Their metric consists of three stages: a multiscale feature extraction network that extracts spatio-temporal features, a hierarchical spatio-temporal fusion network that integrates intermediate feature information, and finally, a quality regression network that predicts the video quality. In \cite{mitra2022multiview}, a completely \ac{bvqa} metric has been proposed. This metric consists of a self-supervised multiview contrastive learning approach, which captures the joint distributions of frame differences with frames and optical flow.  Wu \textit{et al.} proposed a BVQA metric called FAST-VQA~\cite{wu2022fast}, which relies on a new sampling technique, Grid Mini-patch Sampling (GMS). GMS divides a video into spatially non-overlapping grids, randomly selects a mini-patch from each grid, and then assembles and temporally aligns these mini-patches to construct fragments. After sampling, the resultant fragments are fed into the Fragment Attention Network (FANet) to obtain the final video quality score. The same authors have also introduced the DOVER method~\cite{wu2023exploring}, which provides video quality prediction from aesthetic and technical perspectives. Specifically, DOVER metric consists of two branches, each dedicated to focusing on one perspective. Finally, the overall quality score is obtained by a subjectively-inspired fusion of the predictions from the two perspectives. Given the success of the patch-sampling mechanism proposed in FAST-VQA~\cite{wu2022fast}, it was also adopted in~\cite{huang2023xgc}. However, instead of applying the same sampling strategy to all types of videos, in~\cite{huang2023xgc}, the authors proposed to first classify the video into three content types, according to the professionalism of the produced content. Then, based on this classification, different spatial and temporal sampling strategies are applied, thus making it possible  to build a unified \ac{vqa} model.

All these described works only consider visual information, but some recent works have also included audio information in the QA process via a multimodal approach~\cite{min2020study,cao2023subjective}, because the audio information can significantly influence human judgment/perception. Thus, in~\cite{cao2023subjective}, the authors first proposed a novel UGC Audio-Visual Quality Assessment (AVQA) database, which includes UGC audio and video sequences. Then, a deep learning-based approach was proposed, which includes 
 four modules: a visual feature extraction module, an audio feature
extraction module, a temporal pooling module, and finally an audio-visual fusion module that combines the features of the two modalities and provides the final score.

\section{Proposed Double Bi-LSTM Video Quality Assessment Method}
\label{proposed}
Let us consider a video sequence $\mathcal{V}$ as a set of $T$ consecutive frames: $\mathcal{V} = \{\bm{x}_1, \bm{x}_2,  \dots, \bm{x}_T \}$. The problem of UGC-\ac{vqa} is defined as a function $m$ that predicts a quality score $\hat{q}$ from a video sequence $\mathcal{V}$ as follows:  
\begin{equation}
    \hat{q} = m ( \bm{x}_1, \bm{x}_2, \dots, \bm{x}_T).
\end{equation}

To address this problem, we propose a \ac{bvqa} metric called 2BiVQA for double Bi-LSTM Video Quality Assessment. As illustrated in Figure \ref{overall}, the framework of the proposed 2BiVQA is composed of four main modules: features extraction, spatial pooling, temporal pooling, and finally, a quality regression module. These four modules are integrated to form an end-to-end \ac{bvqa} metric. Each of the four modules will be described in detail in the following sections. 
\subsection{Features extraction}
\ac{cnn} features have been shown to correlate well with perceptual judgments~\cite{zhang2018unreasonable} and represent good candidates for human perception-related applications~\cite{gao2017deepsim,yang2020deep,amirshahi2016image,johnson2016perceptual,zhang2018unreasonable}. The performance of \ac{cnn} strongly depends on the number of training samples. However, existing \ac{ugc}-\ac{vqa} datasets are much smaller compared to the typical computer vision datasets with millions of samples.
Thus, it is very difficult to train a deep \ac{cnn} from scratch while achieving competitive quality prediction performance, since the model can be prone to over-fitting problem. Nevertheless, the authors of~\cite{tu2021ugc} showed that well-known deep \ac{cnn} feature descriptors (e.g., ResNet-50~\cite{he2016deep}, VGG-16~\cite{zhang2015accelerating}, etc.) pre-trained on other vision tasks like image classification are transferable to \ac{ugc} \ac{iqa}/\ac{vqa} problems, and they can achieve outstanding performance. In our contribution, we opted for the ResNet-50 pre-trained on ImageNet~\cite{deng2009imagenet} as the backbone model to extract spatial features. Even though several \ac{cnn} models can be used, we obtained the best performance with ResNet-50 (see Tables \ref{table2} and \ref{table3}). In the following, we first introduce the backbone model, and then we detail the feature extraction process.

\vspace{1mm}
\textbf{ResNet-50:}  is a variant of ResNet model that introduced a concept allowing to train ultra-deep neural networks that can include hundreds and even thousands of layers. In fact, theoretically, a deeper neural network is able to learn more complex features, which should lead to better performance. However, in practice, the ultimate effect of adding more and more layers is increasing the training error. This problem is known as the degradation problem~\cite{he2016deep}. Residual blocks introduced in ResNet aim to solve this issue. It includes shortcut connections to perform identity mapping, which allows a deeper model to have no higher training error than its shallower counterpart. 

\vspace{1mm}
\textbf{Features extraction process:} one issue with using pre-trained models is their standard input shape. Two solutions can be envisaged to overcome this constraint, either resizing the input frame or dividing it into several patches. The first technique can affect the perceptual quality or attenuate the intensity of pre-existing artifacts. Therefore, we opted for the second solution, which solves the problem of a standard size input on the one hand, and avoids over-fitting with the limited dataset on another hand. Thus, for each frame $\bm{x}_{i}$ a sliding window is used  to extract $N$  patches $\bm{x}_i^j$, \hbox{$\forall j \in \{1, \dots, N \}, \forall i \in \{1, \dots, T \}$}, with a stride slightly smaller than the patch size  ($224\times224$). Then, these patches are fed into the ResNet-50 model pre-trained on ImageNet for the extraction of spatial features $\bm{y}_i^j$ from the input patch $\bm{x}_i^j$ as follows: 
\begin{equation}
    \bm{y}_i^j =h_\psi ( \bm{x}_i^j), 
\end{equation}
where $h_\psi$ represents the parametric function of the feature extraction model with training parameters $\psi$.

\begin{figure}[t!]
\centering
 \includegraphics[width=0.8\linewidth]{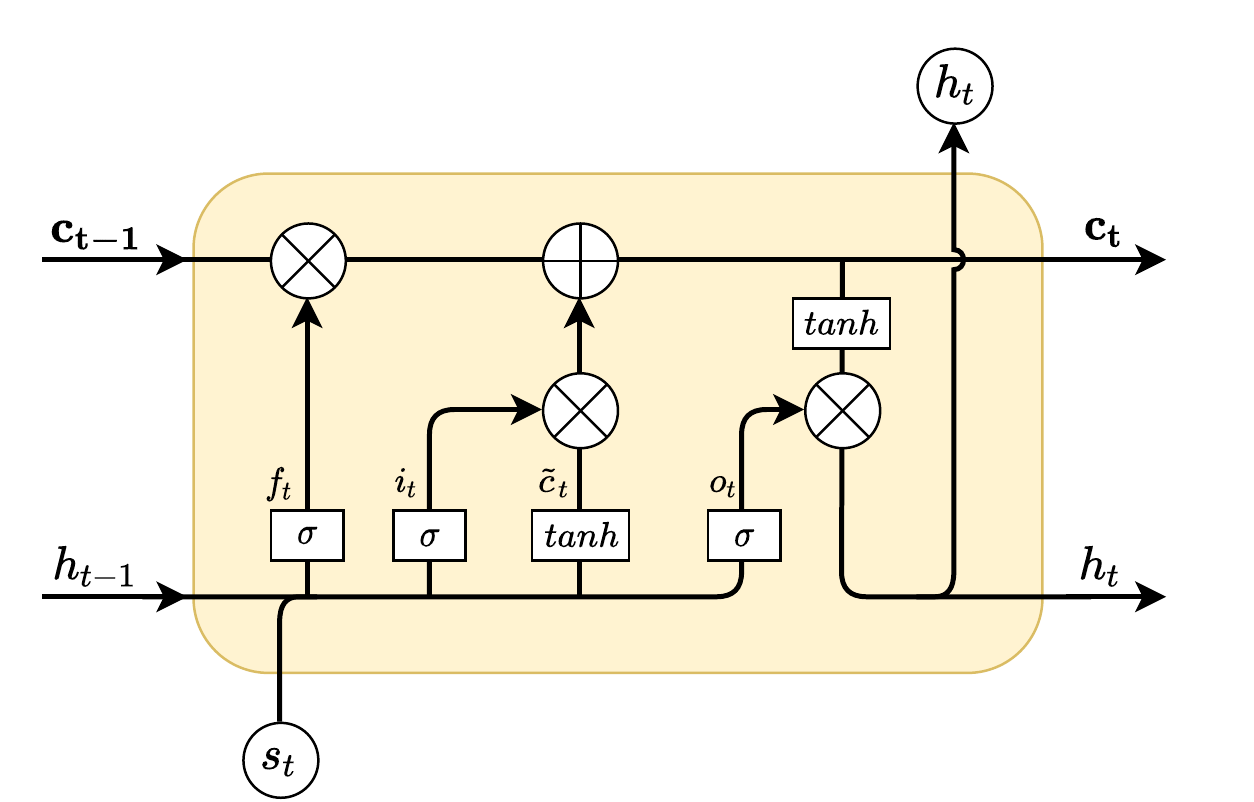}
\caption{The internal structure of the LSTM cell.}
\label{LSTM}\vspace{-4mm}
\end{figure}

\subsection{Spatial pooling}
Once the features have been extracted from each patch, we need to aggregate them into one vector per frame. To do this, it is essential to take into account that \ac{ugc} distortions are not uniformly distributed. In addition, the local distortion visibility is influenced by its surrounding regions, which can either emphasize or mask the perception of distortion. Moreover, the perceptual quality of \ac{hvs} varies over the spatial domain~\cite{watson1997digital}.

Thus, to mimic the \ac{hvs} behavior as well as to account for \ac{ugc} distortion features, unlike processing each patch independently, we consider the entire sequence of patches $(\bm{x}_i^1, \bm{x}_i^2,...,\bm{x}_i^N)^T$ of a frame $i$ at once. To achieve this, we use a sequence model that can efficiently capture the dependencies between the patches of a frame. Specifically, we design our spatial pooling using \ac{bi-lstm} network~\cite{schuster1997bidirectional} as the sequence model, which provides the ability to deal with  dependencies across patches.

In the following, we first explain the internal mechanisms of \ac{lstm}~\cite{hochreiter1997long}, then we introduce \ac{bi-lstm}, and finally, the learning strategy of the proposed spatial pooling module is described.

\vspace{1mm}
\textbf{LSTM} (Long short term memory)~\cite{hochreiter1997long}: is one of the most  popular \acp{rnn},  designed to deal with long time-dependencies. It allows solving the diminishing and exploding gradient problems in  long structures~\cite{hochreiter1998vanishing}. Each \ac{lstm} cell consists of an input gate $i_{t}$, a forget gate $f_{t}$, an output gate $o_{t}$, a candidate cell state $\tilde{c}_{t}$, a cell state $c_{t}$, and a hidden state $h_{t}$, as shown in Figure
~\ref{LSTM}. $i_{t}$ is used to determine the information to store in the current cell state $c_{t}$, while $f_{t}$ determines the thrown away information. $o_{t}$ decides the information to be passed to the current hidden state $h_{t}$, which is computed as follows:
\begin{equation}
\begin{split}
    i_{t} &= \sigma(W^{(i)} \cdot ( h_{t-1} \oplus x_{t} ) + b^{(i)}), \\
    f_{t} &= \sigma(W^{(f)} \cdot ( h_{t-1} \oplus x_{t} ) + b^{(f)}), \\
    o_{t} &= \sigma(W^{(o)} \cdot ( h_{t-1} \oplus x_{t} ) + b^{(o)}),\\
    \tilde{c}_{t} &= \tanh(W^{(c)} \cdot ( h_{t-1} \oplus x_{t} ) + b^{(c)}),\\
    c_{t} &= f_{t} \times c_{t-1} + i_{t} \times \tilde{c}_{t} ,\\
    h_{t} &= o_{t} \times \tanh(c_{t}),
\end{split}
\end{equation}
where $\sigma$ is the sigmoid function and $\oplus$ is the concatenation operator. $+$ and $\times$ are the element-wise addition and product operations, respectively. $W^{(x)}$ and $b^{(x)}$ represent the weight matrix and the bias vector of gate $x$, respectively.

\vspace{1mm}
\textbf{Bi-LSTM} (Bi-directional long short term memory)~\cite{schuster1997bidirectional}: is a stack of two independent \acp{lstm}. This structure allows the network to take both backward and forward information in consideration. It has been proved that \ac{bi-lstm} is far better than regular \ac{lstm} in many fields, like forecasting time series~\cite{siami2019performance}, phoneme classification~\cite{graves2005framewise}, speech recognition~\cite{graves2013hybrid}, etc. However, to the best of our knowledge, \ac{bi-lstm} has not yet been considered in \ac{iqa}/\ac{vqa} problems. 

\begin{figure}[t!]
\centering
 \includegraphics[width=0.8\linewidth]{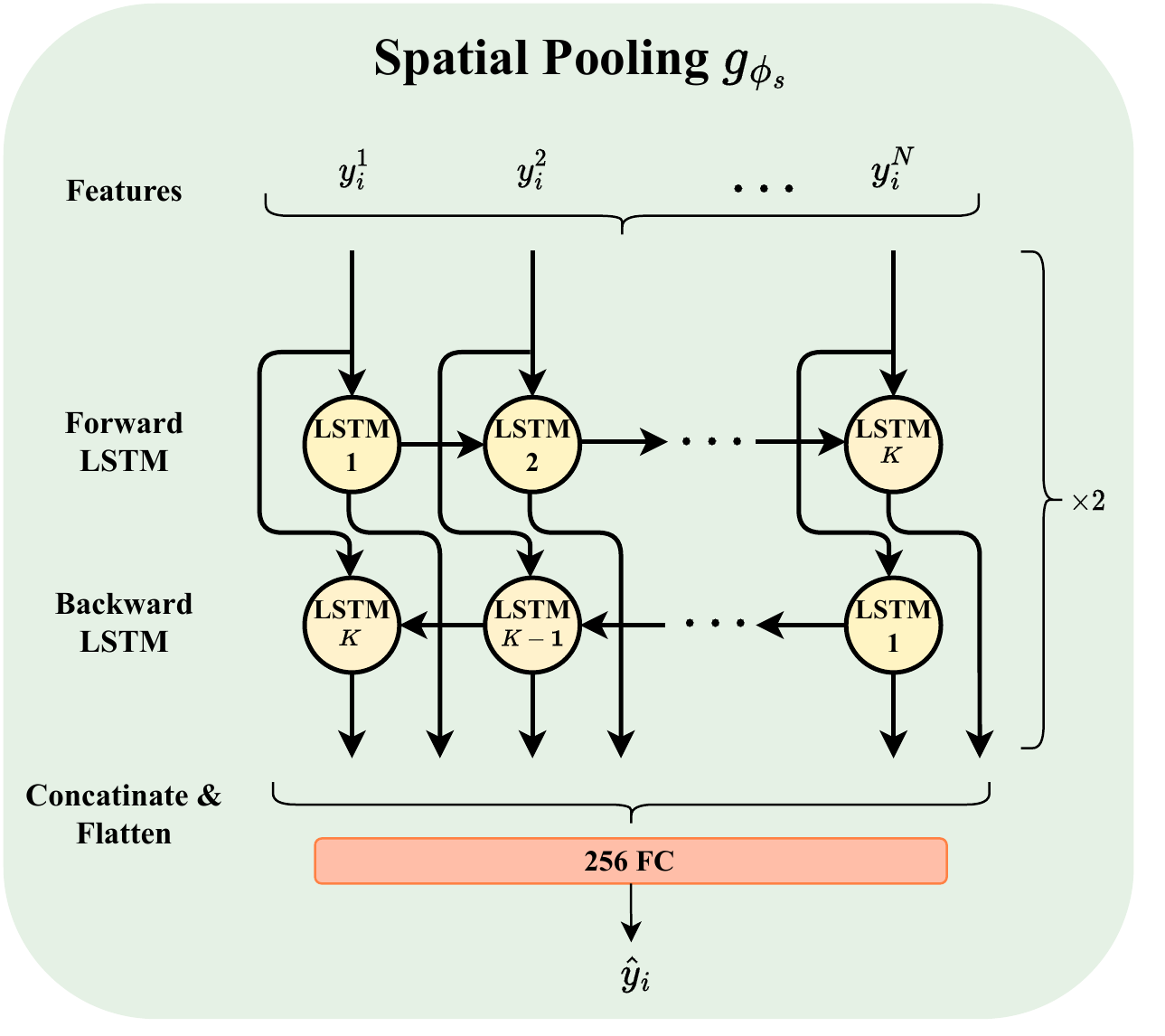}
\caption{The architecture of the spatial pooling module.}
\label{Bi-LSTM}\vspace{-4mm}
\end{figure}

The architecture of the spatial pooling module is shown in Figure~\ref{Bi-LSTM}. The module is composed of two Bi-LSTM layers with $K$ cells each, followed by a \ac{fc} layer with 256 nodes. We have found that using this architecture with $K = 64$ leads to the best results in our experiments. The feature vector $(\bm{y}_i^1, \bm{y}_i^2,  \dots, \bm{y}_i^N)^T$ of a frame $i$ is fed into the spatial pooling module, expressed as:
\begin{equation}
    \hat{\bm{y}}_i = g_{\phi_s} (\bm{y}_i^1 , \bm{y}_i^2, \dots  , \bm{y}_i^N),\hspace*{3mm} \forall i \in \{1, \dots, T \},
\end{equation}
where $g_{\phi_s}$ is the parametric function of the spatial pooling module with the training parameters $\phi_s$.

\vspace{1mm}
\textbf{Pre-training technique}: transfer learning is a powerful machine learning technique. Here, we perform pre-training followed by fine-tuning, which is a widely-used transfer learning paradigm. Pre-training refers to training a
model in a specific source domain $\mathcal{D}_{S}$ with  learning task $\mathcal{T}_{S}$ to initialize its parameters before fine-tuning it for a new learning task $\mathcal{T}_{T}$ in the target domain $\mathcal{D}_{T}$, where a domain $\mathcal{D}$ consists of a feature space $\mathcal{X}$ and a marginal probability distribution $P(X)$, where $X = \left\{\bm{s}_{1}, \ldots, \bm{s}_{n}\right\} \in \mathcal{X}$.



In our approach, the spatial pooling module is trained in this way in two separate stages. It is first pre-trained using the KonIQ-10k  dataset~\cite{hosu2020koniq}, which is a large-scale in-the-wild \ac{iqa} dataset, regardless of the remaining module as an \ac{iqa} model. We assume that previously described complex behaviors and characteristics of \ac{hvs} and \ac{ugc} distortions, respectively, are embedded in the subjective quality dataset. Thus, we aim to transfer the knowledge acquired by the model after training on the KonIQ-10k  dataset, encoded in the weights of the model, to exploit it for the target \ac{ugc}-\ac{vqa} task. Moreover, this pre-training step has the advantage of presenting to the model more diverse content by leveraging the larger authentic \ac{iqa} dataset. 

Finally, the spatial pooling module is fine-tuned using the subjective video quality scores with the rest of the modules in the second
stage.
\subsection{Temporal pooling}
Aggregating quality features of frames into an overall video score is one of the main still unresolved challenges in \ac{vqa}~\cite{seshadrinathan2011temporal,tu2020comparative,park2012video,li2021unified,korhonen2019two}. In fact, the human quality judgment at a late frame can definitely be affected by the previous frames, which is widely known as  the temporal-memory effect~\cite{park2012video,xu2014no,kim2018deep,ninassi2009considering}. According to this temporal behavior of the \ac{hvs}, low-quality frames leave more impressions on the viewer. For instance, if most of the frames are of high quality, and only a few frames are of low quality, humans generally determine that the video is of low quality. Most of the previously developed \ac{vqa} metrics focus much more on the accuracy of quality scores at the frame level, disregarding the impact of adjacent frames, from past and future, on the subjective quality of the current frame.

Therefore, in order to take into account the temporal variation of distortions  as well as the temporal-memory effects of human perception, we propose a novel temporal pooling method using \ac{bi-lstm} network to aggregate frame-level features $(\hat{\bm{y}}_1 , \hat{\bm{y}}_2, \dots , \hat{\bm{y}}_T)$ into a global feature vector $\bar{\bm{y}}$ for the entire video sequence. As described previously, \ac{bi-lstm} networks have the ability to take both backward and forward information into consideration, which makes it possible to capture long-range dependencies between frames like the temporal-memory effect. Similar to spatial pooling, this module is composed of two \ac{bi-lstm} layers with 64 cells each, followed by a \ac{fc} layer with 256 nodes. The temporal pooling module is defined as: 
\begin{equation}
    \bar{\bm{y}} = g_{\phi_t} (\hat{\bm{y}}_1 , \hat{\bm{y}}_2, \dots , \hat{\bm{y}}_T),
\end{equation}
where $g_{\phi_t}$ is the parametric function of the temporal pooling module with the training parameters $\phi_t$.
\subsection{Quality regression}
After extracting quality-aware features and aggregating them into a single vector $\bar{y}$, we need to map these features to the final video quality score $q$. Here, we used one node \ac{fc} as a regression model with a linear activation function. Therefore, we obtain the final video quality score as follows:
\begin{equation}
    \hat{q} = \zeta (\bar{\bm{y}}),
\end{equation}
where $\zeta$ denotes the \ac{fc} layer.

\section{Experimental results}
\label{exp}
\begin{table}[t]
\caption{Summary of the considered \ac{ugc}-\ac{vqa} datasets.}
\label{dataset}
\begin{tabular}{@{}
>{\columncolor[HTML]{FFFFFF}}l 
>{\columncolor[HTML]{FFFFFF}}c 
>{\columncolor[HTML]{FFFFFF}}c 
>{\columncolor[HTML]{FFFFFF}}c 
>{\columncolor[HTML]{FFFFFF}}c 
>{\columncolor[HTML]{FFFFFF}}c @{}}
\toprule 
Database & \#Video & Resolution & Time & Label & Range \\ 

\midrule
KoNViD-1k \cite{hosu2017konstanz}  & 1200  & 540p & 8s & MOS+$\sigma$ & {[}1,5{]} \\
LIVE-VQC \cite{sinno2018large} & 585 & 240p-1080p & 10s & MOS & {[}0,100{]} \\
YouTube-UGC \cite{wang2019youtube} & 1380 & 360p-4k & 20s & MOS+$\sigma$ & {[}1,5{]} \\ \bottomrule 
\end{tabular}
\end{table}

In this section, we first define the experimental setups, including the description of datasets, the evaluation methods  and the implementation details. Then, we present the results of ablation studies, the comparison with \ac{sota}, the generalization capability, and finally, the complexity evaluation.
\subsection{Datasets}
To train, fine-tune and test the proposed model, three \ac{ugc}-\ac{vqa} datasets were considered, including KoNViD-1K~\cite{hosu2017konstanz}, LIVE-VQC~\cite{sinno2018large} and YouTube-UGC~\cite{wang2019youtube}.
The features of these three datasets are summarized in Table~\ref{dataset}. 
For YouTube-UGC, we excluded 57 grayscale videos for a fair comparison as in~\cite{tu2021ugc}. We also used these three datasets to create a fourth dataset, which is the union of them after \ac{mos} calibration using the \ac{insla}~\cite{pinson2003objective, tu2021ugc}: 
\begin{align}
    q' &= 5 - 4 \times \left ((5 - q)/4 \times 1.1241 - 0.0993 \right ),
\label{konvid_eq}\\
    q' &= 5 - 4 \times \left ( (100 - q)/100 \times 0.7132 + 0.0253 \right ),
\label{live_eq}
\end{align}
where Eqs. \eqref{konvid_eq} and \eqref{live_eq} are used for calibrating KoNViD-1K and LIVE-VQC, respectively, while YouTube-UGC is selected as the anchor dataset. $q'$ and $q$ denote the adjusted and the original scores, respectively. The formed dataset is referred to in the following as All-Combined.

In addition, the KonIQ-10k  \ac{iqa} dataset~\cite{hosu2020koniq} is used to train the spatial pooling module separately. This dataset contains 10,073 in-the-wild images with a  resolution of 1024$\times$768 pixels.
\subsection{Evaluation methods}
Since there is no defined size of the training (or test) set  for KoNViD-1k, LIVE-VQA, and YouTube-UGC, and as convened, we randomly split each dataset into two non-overlapping subsets, 80\% for training and 20\% for testing. We performed $k$ fold iterations, and the median performance on the test sets is reported.

We considered four standard measures to assess the performance of the proposed model, including \ac{srocc} and \ac{krcc}, which are  prediction monotonicity measures, and \ac{plcc} and \ac{rmse}, which are  prediction accuracy measures. Before calculating \ac{plcc} and \ac{rmse}, we performed a nonlinear four-parametric logistic regression to match the predicted score to the subject score as follows:
\begin{equation}
    L_{g}(x) = \beta + \frac{\alpha - \beta}{1 + \exp({-x + \gamma /  |\delta|})}.
\end{equation}

\subsection{Training process}
The training is conducted in two steps: first, the spatial pooling module is pre-trained, then the spatial and temporal pooling modules are trained end-to-end.

\begin{figure}[t!]
\centering
 \includegraphics[width=0.88\linewidth]{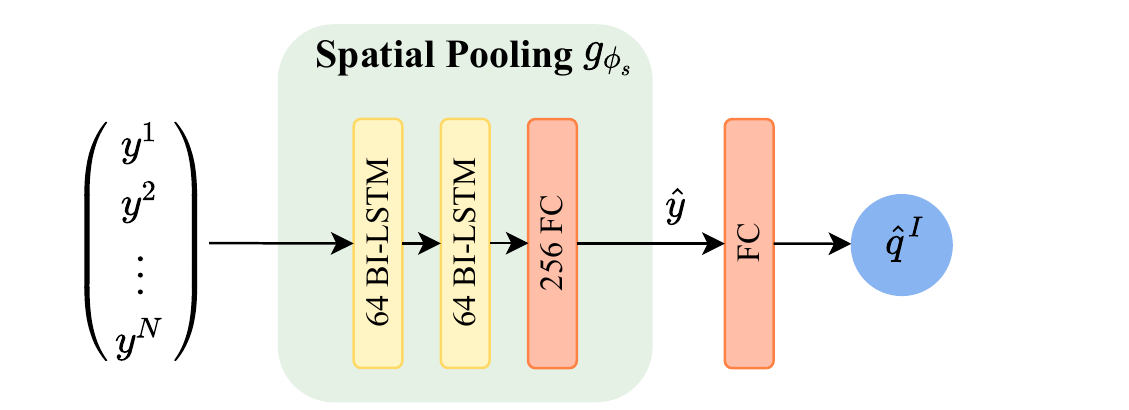}
\caption{The architecture of the spatial pooling module for pre-training.}
\label{sp_train}
\end{figure}

For the pre-training step, each image from KonIQ-10k dataset is divided into $N$ patches. Then, the features extracted from the patches $({y}^1, {y}^2, \ldots ,{y}^N)^T$ are fed into the spatial pooling module for training. A \ac{fc} layer with an identity activation function is added as a regressor head to predict the image quality score $\hat{q}^I$, as illustrated in Figure \ref{sp_train}.
\begin{table}[!]
\caption{Performance of the ablation study for the spatial pooling module on the KonIQ-10k dataset. Bold entries indicate the top three performing methods, while the best method is underlined.}
\adjustbox{max width=0.49\textwidth}{%
\label{table2}
\begin{tabular}{@{}
>{}c 
>{}c 
>{}c
>{}c 
>{}c 
>{}c 
>{}c @{}}
\toprule 
\multirow{2}{*}{\rotatebox{90}{Model}}     & \#Features  & 
Spatial & \multirow{2}{*}{SROCC $\uparrow$}        & \multirow{2}{*}{PLCC $\uparrow$}        & \multirow{2}{*}{KRCC $\uparrow$}        & \multirow{2}{*}{RMSE $\downarrow$}        \\
 & per patch &pooling & & & & \\
\midrule
\cellcolor[HTML]{FFFFFF}                                  & \cellcolor[HTML]{FFFFFF}                                       & Concatenate                                  & 0.821      & 0.829     & 0.628     & 0.378     \\
\cellcolor[HTML]{FFFFFF}                                  & \cellcolor[HTML]{FFFFFF}                                       & Mean                                         & 0.818      & 0.829     & 0.622     & 0.381     \\
\cellcolor[HTML]{FFFFFF}                                  & \cellcolor[HTML]{FFFFFF}                                       & LSTM                                         & 0.875      & 0.894     & 0.689     & 0.249     \\
\multirow{-4}{*}{\cellcolor[HTML]{FFFFFF}\rotatebox{90}{VGG16}}            & \multirow{-4}{*}{\cellcolor[HTML]{FFFFFF}512}                  & Bi-LSTM                                      & 0.876      & 0.900     & 0.693     & 0.242     \\ \midrule
\cellcolor[HTML]{FFFFFF}                                  & \cellcolor[HTML]{FFFFFF}                                       & Concatenate                                  & 0.833      & 0.850     & 0.639     & 0.293     \\
\cellcolor[HTML]{FFFFFF}                                  & \cellcolor[HTML]{FFFFFF}                                       & Mean                                         & 0.825      & 0.839     & 0.623     & 0.373     \\
\cellcolor[HTML]{FFFFFF}                                  & \cellcolor[HTML]{FFFFFF}                                       & LSTM                                         & 0.871      & 0.898     & 0.691     & 0.245     \\
\multirow{-4}{*}{\cellcolor[HTML]{FFFFFF}\rotatebox{90}{Densenet169}}     & \multirow{-4}{*}{\cellcolor[HTML]{FFFFFF}1664}                 & Bi-LSTM                                      & \textbf{0.878}      & \textbf{0.902}     & \textbf{0.698}     & \textbf{0.240}     \\ \midrule
\cellcolor[HTML]{FFFFFF}                                  & \cellcolor[HTML]{FFFFFF}                                       & Concatenate                                  & 0.856      & 0.871     & 0.669     & 0.275     \\
\cellcolor[HTML]{FFFFFF}                                  & \cellcolor[HTML]{FFFFFF}                                       & Mean                                         & 0.856      & 0.858     & 0.664     & 0.275     \\
\cellcolor[HTML]{FFFFFF}                                  & \cellcolor[HTML]{FFFFFF}                                       & LSTM                                         & \textbf{0.906}      & \textbf{0.924}     & \textbf{0.737}     & \textbf{0.214}     \\
\multirow{-4}{*}{\cellcolor[HTML]{FFFFFF}\rotatebox{90}{ResNet50}}        & \multirow{-4}{*}{\cellcolor[HTML]{FFFFFF}2048}                 & Bi-LSTM                                      & {\ul \textbf{0.910}}      & {\ul \textbf{0.925}}     & {\ul \textbf{0.742}}     & {\ul \textbf{0.213}}     \\ \midrule
\cellcolor[HTML]{FFFFFF}                                  & \cellcolor[HTML]{FFFFFF}                                       & Concatenate                                  & 0.763      & 0.741     & 0.570     & 0.414     \\
\cellcolor[HTML]{FFFFFF}                                  & \cellcolor[HTML]{FFFFFF}                                       & Mean                                         & 0.791      & 0.816     & 0.593     & 0.391     \\
\cellcolor[HTML]{FFFFFF}                                  & \cellcolor[HTML]{FFFFFF}                                       & LSTM                                         & 0.851      & 0.873     & 0.664     & 0.272     \\
\multirow{-4}{*}{\cellcolor[HTML]{FFFFFF}\rotatebox{90}{EﬀicientNetB7}} & \multirow{-4}{*}{\cellcolor[HTML]{FFFFFF}2560}                 & Bi-LSTM                                      & 0.853      & 0.878     & 0.668     & 0.266     \\ 
\bottomrule
\end{tabular}

}
\vspace{1ex}

\vspace{-5mm}
\end{table}

The pre-training process is performed with 200 epochs using Adam optimizer~\cite{DBLP:journals/corr/KingmaB14} with an initial learning rate of $1e-4$, batch size of 16 and the \ac{mse} as loss function $\ell_2$: 

\begin{equation}
    \ell_2({q}^I,\hat{q}^I) = \frac{1}{L} \sum_{l=1}^{L} \left ({q}_l^I - \hat{q}_l^I\right )^2,
\end{equation}
where ${q}^I, {\hat{q}}^I$ and $L$ represent the ground truth, the predicted image quality score and the batch size, respectively. 

The second stage of the training process is to map the frame-level feature vectors into the global quality score. For this purpose, the spatial pooling module is fine-tuned at the same time as the temporal pooling module on the target \ac{ugc}-\ac{vqa} datasets. This is done using the same hyper-parameters as the pre-training step: 200 epochs with an initial learning rate of $1e-4$ and \ac{mse} as a loss function $\ell_2(q,\hat{q})$ computed between $q$ and $\hat{q}$, which represent the ground truth and the predicted video quality scores, respectively.


Note that during the pre-training and fine-tuning steps, the weights of the backbone model are frozen. To support the principle of reproducible research and fair comparison, an implementation of the 2BiVQA metric is made publicly available for the research community\footnote{\href{https://github.com/atelili/2BiVQA}{https://github.com/atelili/2BiVQA}}.

\begin{table*}[t!]
\centering 
\caption{Performance of the ablation study on the KoNViD-1k dataset. Each entry is presented as spatial pooling  without/with pre-training on the KonIQ-10k dataset. Bold entries indicate the top three performing methods, while the best method is underlined.}
\label{table3}
\begin{tabular}{@{}
>{\columncolor[HTML]{FFFFFF}}l 
>{\columncolor[HTML]{FFFFFF}}l 
>{\columncolor[HTML]{FFFFFF}}c 
>{\columncolor[HTML]{FFFFFF}}c 
>{\columncolor[HTML]{FFFFFF}}c 
>{\columncolor[HTML]{FFFFFF}}c 
>{\columncolor[HTML]{FFFFFF}}c 
>{\columncolor[HTML]{FFFFFF}}c 
>{\columncolor[HTML]{FFFFFF}}c @{}}
\toprule
\cellcolor[HTML]{FFFFFF}{Model}                &  & \cellcolor[HTML]{FFFFFF} Temporal pooling    & \cellcolor[HTML]{FFFFFF} & SROCC $\uparrow$                 & PLCC $\uparrow$                  & KRCC $\uparrow$                 & RMSE $\downarrow$                  & \cellcolor[HTML]{FFFFFF}\#parameters \\ \midrule
\cellcolor[HTML]{FFFFFF}                                   &  & Mean      &                                            & 0.776 / 0.771               & 0.773  / 0.767               & 0.588   / 0.575              & 0.424 / 0.451                & 1,213,953                                               \\ 

\cellcolor[HTML]{FFFFFF}                                  &  & Harmonic  &                                            & 0.776 / 0.773                & 0.774 / 0.769                & 0.588   / 0.576               & 0.424 / 0.453                & 1,213,953                                               \\ 
\cellcolor[HTML]{FFFFFF}                                   &  & Geometric &                                            & 0.777 / 0.772                 & 0.774 / 0.768                & 0.588 / 0.576                & 0.424 / 0.452                & 1,213,953                                               \\
\cellcolor[HTML]{FFFFFF}                                   &  & LSTM      &                                            & 0.790 / 0.809               & 0.799 / 0.829                 & 0.594 / 0.613                & 0.461 / 0.383               & 1,820,929                                               \\      
\multirow{-5}{*}{\cellcolor[HTML]{FFFFFF}{VGG16}}                                   &  & Bi-LSTM   &                                            & 0.797 / 0.819                & 0.806 / 0.833                & 0.606 / 0.622               & 0.458  / 0.380              & 2,529,281                                               \\ \midrule
\cellcolor[HTML]{FFFFFF}                                  &  & Mean      &                                            & 0.799 / 0.815                & 0.806 / 0.811                & 0.601 / 0.624               & 0.414 / 0.404                & 1,803,777                                               \\ 
\cellcolor[HTML]{FFFFFF}                                   &  & Harmonic  &                                            & 0.800 / 0.816                & 0.807 / 0.812                & 0.603  / 0.624              & 0.415 /  0.403               & 1,803,777                                               \\
\cellcolor[HTML]{FFFFFF}                                  &  & Geometric &                                            & 0.799 / 0.815               & 0.807  / 0.811                & 0.602    / 0.624            & 0.415 / 0.404               & 1,803,777                                               \\
\cellcolor[HTML]{FFFFFF}                                  &  & LSTM      &                                            & 0.805 / 0.825               & 0.821 / \textbf{0.839}               & 0.618 / 0.623                & 0.392  / \textbf{0.373}              & 2,410,753                                               \\
\multirow{-5}{*}{\cellcolor[HTML]{FFFFFF}{DenseNet169}}                                    &  & Bi-LSTM   &                                            & 0.810 / 0.825                & 0.824 / {\ul \textbf{0.842}}             & 0.621 / 0.626                & 0.385 / \textbf{0.370}               & 3,050,241                                                                   \\ \midrule
\cellcolor[HTML]{FFFFFF}                                   &  & Mean      &                                            & 0.795 / 0.829               & 0.802 / 0.816                & 0.601 / 0.631               & 0.411 / 0.398               & 2,000,385                                               \\
\cellcolor[HTML]{FFFFFF}                                   &  & Harmonic  &                                            & 0.796 / 0.828                 & 0.802 / 0.815                & 0.603 / 0.631                & 0.409 / 0.400                & 2,000,385                                               \\
\cellcolor[HTML]{FFFFFF}                                   &  & Geometric &                                            & 0.795 / \textbf{0.829}                & 0.802 / 0.815                & 0.602 / 0.631                & 0.410 / 0.399                & 2,000,385                                               \\
\cellcolor[HTML]{FFFFFF}                                   &  & LSTM      &                                            & 0.825 / 0.827               & 0.821 / 0.819               & 0.625 / \textbf{0.633}                & 0.394  / 0.384              & 2,607,361                                               \\
\multirow{-5}{*}{\cellcolor[HTML]{FFFFFF}{ResNet50}}                                    &  & Bi-LSTM   &                                            & \textbf{0.830} / {\ul \textbf{0.846}}       & 0.820 / \textbf{0.840}                & \textbf{0.634} / {\ul \textbf{0.652}}        & 0.382 / {\ul \textbf{0.362}}                & 3,312,385                                               \\ \midrule
\cellcolor[HTML]{FFFFFF}                                  &  & Mean      &                                            & 0.746 / 0.782                 & 0.766 / 0.780                & 0.557 / 0.594                & 0.458 / 0.432                & 2,262,529                                               \\
\cellcolor[HTML]{FFFFFF}                                   &  & Harmonic  &                                            & 0.749 / 0.785                & 0.770   / 0.783             & 0.559 / 0.597                & 0.457 / 0.432                & 2,262,529                                               \\
\cellcolor[HTML]{FFFFFF}                                   &  & Geometric &                                            & 0.747 / 0.784               & 0.768  / 0.782              & 0.558 / 0.596               & 0.457 / 0.432                & 2,262,529                                               \\
\cellcolor[HTML]{FFFFFF}                                   &  & LSTM      &                                            & 0.752 / 0.800               & 0.773  / 0.809              & 0.561 / 0.602               & 0.451    / 0.398            & 2,869,505                                               \\
\multirow{-5}{*}{\cellcolor[HTML]{FFFFFF}{EfficientNetB7}}                                   &  & Bi-LSTM   &                                            & 0.759 / 0.801                 & 0.776 / 0.814                & 0.567 / 0.605                & 0.448 / 0.398                & 3,508,993                                                                      \\ \bottomrule 
\end{tabular}
\end{table*}

\begin{table*}[htbp]
\small
\centering
\caption{Performance comparison of evaluated BVQA models on the four \ac{ugc}-\ac{vqa} datasets. The underlined and boldfaced entries indicate the best and top three performers on each dataset for each performance measure, respectively.}
\label{table4}
\adjustbox{max width=0.87\textwidth}{
\begin{tabular}{@{}
>{\columncolor[HTML]{FFFFFF}}l| 
>{\columncolor[HTML]{FFFFFF}}c 
>{\columncolor[HTML]{FFFFFF}}c 
>{\columncolor[HTML]{FFFFFF}}c 
>{\columncolor[HTML]{FFFFFF}}c 
>{\columncolor[HTML]{FFFFFF}}l 
>{\columncolor[HTML]{FFFFFF}}c 
>{\columncolor[HTML]{FFFFFF}}c 
>{\columncolor[HTML]{FFFFFF}}c 
>{\columncolor[HTML]{FFFFFF}}c @{}}
\toprule
Dataset & \multicolumn{4}{c}{\cellcolor[HTML]{FFFFFF}KonViD-1k} &  & \multicolumn{4}{c}{\cellcolor[HTML]{FFFFFF}LIVE-VQC} \\ \cmidrule(lr){2-5} \cmidrule(l){7-10} 
Model  & SROCC $\uparrow$ & PLCC $\uparrow$ & KRCC $\uparrow$ & RMSE $\downarrow$ &  & SROCC $\uparrow$ & PLCC $\uparrow$ & KRCC $\uparrow$ & RMSE$\downarrow$ \\ \midrule
BRISQUE \cite{mittal2012no} & 0.656 & 0.657 & 0.476 & 0.481 &  & 0.592 & 0.638 & 0.416 & 13.100 \\ 
NIQE \cite{mittal2012making} & 0.541 & 0.553 & 0.379 & 0.533 & \multicolumn{1}{c}{\cellcolor[HTML]{FFFFFF}} & 0.595 & 0.628 & 0.425 & 13.110 \\ 
ILNIQE \cite{zhang2015feature} & 0.526 & 0.540 & 0.369 & 0.540 & \multicolumn{1}{c}{\cellcolor[HTML]{FFFFFF}} & 0.503 & 0.543 & 0.355 & 14.148 \\ 
StairIQA \cite{sun2023blind} & 0.767 & 0.778 & 0.570 & 0.881 & \multicolumn{1}{c}{\cellcolor[HTML]{FFFFFF}} & 0.740 & 0.787 & 0.547 & 10.597 \\ 
BMPRI \cite{min2018blind} & 0.519 & 0.522 & 0.354 & 0.661 &  & 0.502 & 0.586 & 0.350 & 13.558 \\
HIGRADE \cite{kundu2017no} & 0.720 & 0.726 & 0.531 & 0.439 &  & 0.610 & 0.633 & 0.439 & 13.027 \\ 
FRIQUEE \cite{ghadiyaram2017perceptual} & 0.747 & 0.748 & 0.550 & 0.425 &  & 0.657 & 0.700 & 0.477 & 12.198 \\ 
CORNIA \cite{ye2012unsupervised} & 0.716 & 0.713 & 0.523 & 0.448 &  & 0.671 & 0.718 & 0.484 & 11.832 \\ 
HOSA \cite{xu2016blind} & 0.765 & 0.766 & 0.569 & 0.414 &  & 0.687 & 0.741 & 0.503 & 11.353 \\ 
VGG-19 \cite{simonyan2014very} & 0.774 & 0.784 & 0.584 & 0.395 &  & 0.656 & 0.716 & 0.472 & 11.783 \\ 
ResNet-50 \cite{he2016deep} & 0.801 & 0.810 & 0.610 & 0.374 &  & 0.663 & 0.720 & 0.478 & 11.591 \\
KonCept512 \cite{hosu2020koniq} & 0.734 & 0.748 & 0.542 & 0.426 &  & 0.664 & 0.727 & 0.479 & 11.626 \\
PaQ-2-PiQ \cite{ying2020patches} & 0.613 & 0.601 & 0.433 & 0.514 &  & 0.643 & 0.668 & 0.456 & 12.619 \\
V-BLIINDS \cite{saad2014blind} & 0.710 & 0.703 & 0.518 & 0.459 &  & 0.693 & 0.717 & 0.507 & 11.765 \\
TLVQM \cite{korhonen2019two} & 0.772 & 0.768 & 0.577 & 0.410 &  & {\ul \textbf{0.798}} & \textbf{0.802} & \textbf{0.608} & \textbf{10.145} \\
VIDEVAL \cite{tu2021ugc} & 0.783 & 0.780 & 0.584 & 0.402 &  & 0.752 & 0.751 & 0.563 & 11.100 \\
RAPIQUE \cite{tu2021rapique} & \textbf{0.807} & \textbf{0.815} & \textbf{0.618} & \textbf{0.364} &  & 0.741 & 0.765 & 0.557 & 10.665 \\
FAST-VQA \cite{wu2022fast} & {\ul\textbf{0.846}} & {\ul\textbf{0.854}} & {\ul\textbf{0.638}} & {\ul\textbf{0.337}} &  & \textbf{0.792} & {\ul\textbf{0.844}} & {\ul\textbf{0.633}} & {\ul\textbf{9.904}} \\
2BiVQA &  \textbf{0.815} & \textbf{0.835} &  \textbf{0.629} & \textbf{0.352} &  & \textbf{0.761} &  \textbf{0.832} &  \textbf{0.621} & \textbf{9.979} \\ \midrule
Dataset & \multicolumn{4}{c}{\cellcolor[HTML]{FFFFFF}YouTube-UGC} & \multicolumn{1}{c}{\cellcolor[HTML]{FFFFFF}} & \multicolumn{4}{c}{\cellcolor[HTML]{FFFFFF}All-Combined} \\ \cmidrule(lr){2-5} \cmidrule(l){7-10} 
Model & SROCC $\uparrow$ & PLCC $\uparrow$ & KRCC $\uparrow$ & RMSE $\downarrow$ &  & SROCC $\uparrow$ & PLCC $\uparrow$ & KRCC $\uparrow$ & RMSE $\downarrow$ \\ \midrule
BRISQUE \cite{mittal2012no} & 0.382 & 0.395 & 0.263 & 0.591 &  & 0.569 & 0.586 & 0.403 & 0.561 \\
NIQE \cite{mittal2012making} & 0.237 & 0.277 & 0.160 & 0.617 & \multicolumn{1}{c}{\cellcolor[HTML]{FFFFFF}} & 0.462 & 0.477 & 0.322 & 0.611 \\
ILNIQE \cite{zhang2015feature} & 0.291 & 0.330 & 0.198 & 0.605 & \multicolumn{1}{c}{\cellcolor[HTML]{FFFFFF}} & 0.459 & 0.474 & 0.321 & 0.611 \\
StairIQA \cite{sun2023blind} & 0.753 & 0.744 & 0.558 & 0.597 & \multicolumn{1}{c}{\cellcolor[HTML]{FFFFFF}} & 0.777 & 0.780 & 0.586 & 0.480 \\
BMPRI \cite{min2018blind} & 0.295 & 0.372 & 0.199 & 0.636 &  & 0.505 & 0.525 & 0.351 & 0.678 \\
HIGRADE \cite{kundu2017no} & 0.737 & 0.721 & 0.547 & 0.447 &  & 0.739 & 0.736 & 0.547 & 0.467 \\
FRIQUEE \cite{ghadiyaram2017perceptual} & 0.765 & 0.757 & 0.568 & 0.416 &  & 0.756 & 0.755 & 0.565 & 0.454 \\
CORNIA \cite{ye2012unsupervised} & {\color[HTML]{24292E} 0.597} & 0.605 & 0.421 & 0.513 &  & {\color[HTML]{24292E} 0.676} & 0.697 & 0.484 & 0.494 \\
HOSA \cite{xu2016blind} & 0.602 & 0.604 & 0.425 & 0.513 &  & {\color[HTML]{24292E} 0.695} & 0.708 & 0.503 & 0.489 \\
VGG-19 \cite{simonyan2014very} & 0.702 & 0.699 & 0.509 & 0.456 &  & 0.732 & 0.748 & 0.539 & 0.461 \\
ResNet-50 \cite{he2016deep} & {\color[HTML]{24292E} 0.718} & 0.709 & 0.522 & 0.453 &  & {\color[HTML]{24292E} 0.755} & 0.774 & 0.561 & 0.438 \\
KonCept512 \cite{hosu2020koniq} & 0.587 & 0.594 & 0.410 & 0.513 &  & {\color[HTML]{24292E} 0.660} & 0.676 & 0.475 & 0.509 \\
PaQ-2-PiQ \cite{ying2020patches} & 0.265 & 0.293 & 0.177 & 0.615 &  & 0.472 & 0.482 & 0.324 & 0.608 \\
V-BLIINDS \cite{saad2014blind} & 0.559 & 0.555 & 0.389 & 0.535 &  & 0.654 & 0.6599 & 0.473 & 0.520 \\
TLVQM \cite{korhonen2019two} & 0.669 & 0.659 & 0.481 & 0.484 &  & 0.727 & 0.734 & 0.534 & 0.470 \\
VIDEVAL \cite{tu2021ugc} &  \textbf{0.778} & \textbf{0.773} &  \textbf{0.583} &  \textbf{0.404} &  & 0.796 & 0.793 & 0.603 & 0.426 \\
RAPIQUE \cite{tu2021rapique} & 0.761 & 0.762 & 0.561 & 0.406 &  & {\ul \textbf{0.808}} & {\ul \textbf{0.818}} & {\ul \textbf{0.614}} & {\ul \textbf{0.407}} \\
FAST-VQA \cite{wu2022fast} & {\ul\textbf{0.811}} & {\ul\textbf{0.817}} & {\ul\textbf{0.619}} & {\ul\textbf{0.386}} &  & \textbf{0.804} & \textbf{0.800} & \textbf{0.606} & \textbf{0.453} \\
2BiVQA & \textbf{0.771} & \textbf{0.790} & \textbf{0.581} & \textbf{0.404} &  & \textbf{0.800} & \textbf{0.794} & \textbf{0.608} & \textbf{0.421} \\ \bottomrule
\end{tabular}%
}
\end{table*}

\subsection{Ablation studies} 
To justify the choice and highlight the contribution of each component in the proposed model, we conducted ablation studies on the following aspects. The first study aims to select the best backbone model to extract reliable perceptual features. We considered four well-known deep \ac{cnn} models: VGG16~\cite{zhang2015accelerating},  Densenet169~\cite{huang2017densely}, ResNet-50~\cite{he2016deep} and EfficientNetB7~\cite{tan2019efficientnet}. Also, we tested four spatial pooling methods: simple concatenation, arithmetic mean, \ac{lstm}, and \ac{bi-lstm}. The result of this first study is reported in Table \ref{table2}, where some interesting observations can be made. First, ResNet-50 is able to extract the most significant perceptual features and achieves better performance than the other \ac{cnn} models. Second, \ac{rnn} models (\ac{lstm} and \ac{bi-lstm}) are the pooling methods that obtained the highest correlation scores allowing a 6.09\% improvement in terms of \ac{srocc} compared to the classical pooling methods, including concatenation and arithmetic mean. Finally, this study shows that the best combination for the \ac{iqa} is ResNet-50 as a features extraction model and \ac{bi-lstm} as a spatial pooling method.

In the second study, we investigated the effect of pre-training the spatial pooling module, and we also tested several temporal pooling methods, including arithmetic mean, harmonic mean, geometric mean, \ac{lstm}, and \ac{bi-lstm}. We depict the results of this second study in Table \ref{table3}. It is important to note that this study is conducted on KoNViD-1K dataset with a randomly 80\%-20\% split over only one iteration to avoid a huge training time. The results show that pre-training the spatial pooling module on KonIQ-10k  dataset significantly improves the prediction performance, for instance in terms of \ac{srocc} by 3.04$\%$. Moreover, the results indicate that \ac{bi-lstm} is the best-performing temporal pooling method, showing its effectiveness in capturing long-range dependencies between frames.

\begin{figure*}[!t]
\subfloat[VIDEVAL]{\includegraphics[width=\linewidth]{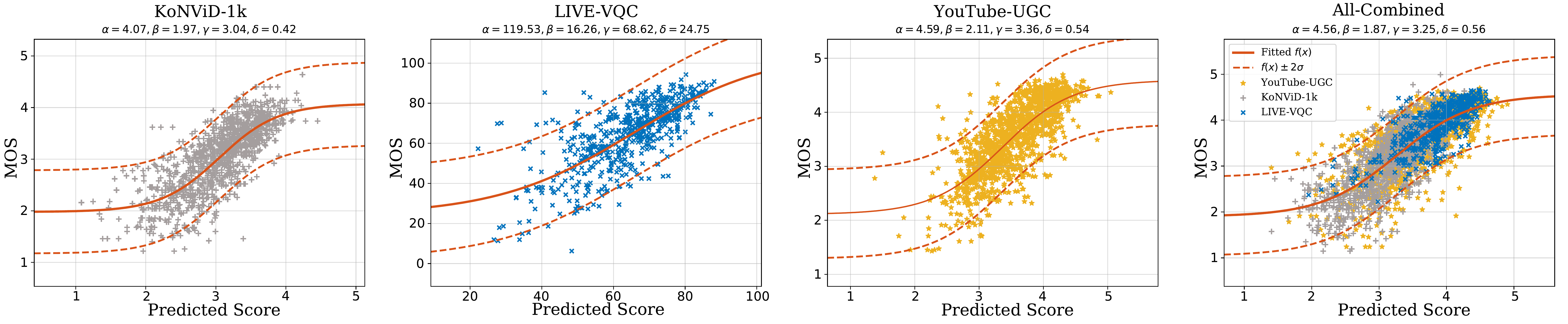}\label{fig_VIDEVA}}
\hfil
\subfloat[RAPIQUE]{\includegraphics[width=\linewidth]{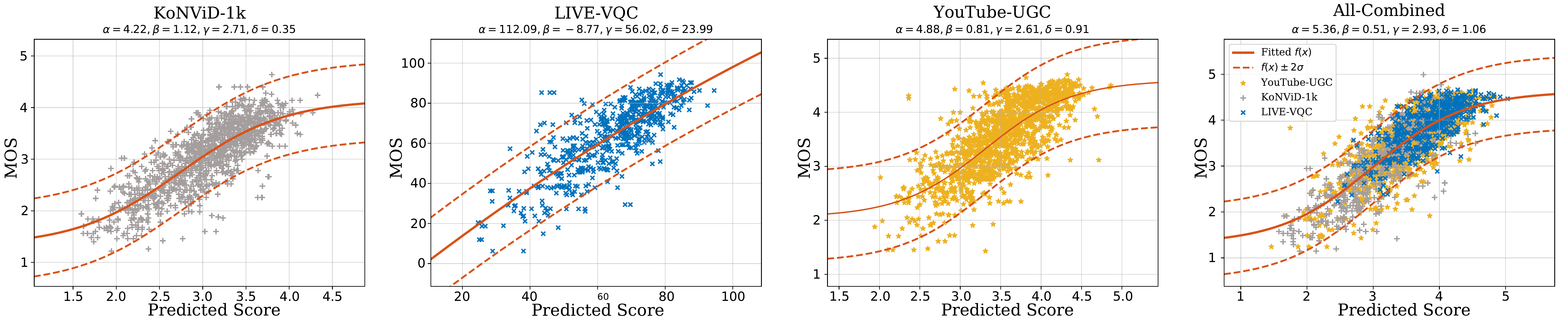}\label{fig_rapique}}
\hfil
\subfloat[2BiVQA]{\includegraphics[width=\linewidth]{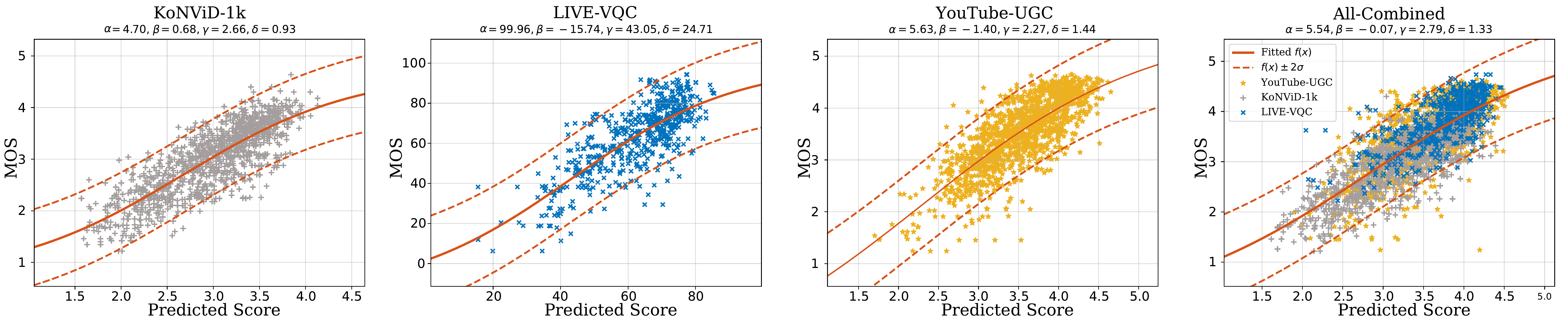}\label{fig_2BiVQA}}
\hfil
\caption{Scatter plots and nonlinear logistic fitted curves of VIDEVAL, RAPIQUE, and 2BiVQA models versus MOS  using k-fold cross-validation on KoNViD-1k, LIVE-VQC, YouTube-UGC, and All-Combined datasets. The logistic model coefficients are given for the three objective metrics tested on the four datasets.}
\label{scatter}
\end{figure*}
\begin{table}[]
\centering
\caption{Comparison of 2BiVQA and FAST-VQA characteristics during training.}
\label{com}
\adjustbox{max width=1\textwidth}{%
\begin{tabular}{@{}lccc@{}}
\toprule
Model   & \# parameters & Training time & Energy    \\ \midrule
FAST-VQA & 28,127,901   & 7920 s        & 384.92 Wh \\
2BiVQA   & 3,246,849    & 580 s         & 11.21 Wh  \\ \bottomrule
\end{tabular}%
}
\end{table}

\subsection{Performance evaluation and comparison} \label{Main Evaluation Results}
To assess the performance of the proposed 2BiVQA metric, we compared it with ten \ac{biqa} models (BRISQUE~\cite{mittal2012no}, NIQE~\cite{mittal2012making}, ILNIQE~\cite{zhang2015feature}, BMPRI~\cite{min2018blind}, StairIQA~\cite{sun2023blind},  HIGRADE~\cite{kundu2017no}, CORNIA~\cite{ye2012unsupervised},  HOSA~\cite{xu2016blind},  FRIQUEE~\cite{ghadiyaram2017perceptual}, PaQ-2-PiQ~\cite{ying2020patches} and KonCept512~\cite{hosu2020koniq}), and five \ac{bvqa} models (V-BLIINDS~\cite{saad2014blind}, FAST-VQA~\cite{wu2022fast}, \ac{tlvqm}~\cite{korhonen2019two}, \ac{videval}~\cite{tu2021ugc} and RAPIQUE~\cite{tu2021rapique}). In addition, two deep \ac{cnn} models (VGG-19~\cite{simonyan2014very} and ResNet-50~\cite{he2016deep}) using transfer learning were benchmarked. Among these methods, NIQE and ILNIQE  are completely blind because they don't require any training. The rest of the methods were trained and tested under the same conditions as our proposed model. For  VGG-19 and ResNet-50 models, the frame-level scores are obtained using two \ac{fc} layers with 256 and 1 nodes, respectively. For all considered \ac{biqa} models, we extend them for \ac{vqa} by averaging the separate frame quality scores to obtain the overall video quality score.

Table~\ref{table4} shows the performance of these methods on the four considered datasets. We can notice that most of the \ac{biqa} metrics, except those \ac{cnn}-based, provide low performance, which indicates that the temporal-related features are substantial for \ac{vqa}, and using a simple average pooling is not sufficient to achieve high performance. We can also observe that \ac{cnn}-based \ac{biqa} approaches, i.e., VGG-19 and ResNet-50, perform well on larger datasets (KoNViD-1k, YouTube-UGC and All-Combined), showing the superiority of the data-driven deep-learning approaches over handcrafted feature-based ones when trained with sufficient dataset size.

On KoNViD-1K, \ac{bvqa} methods generally provide acceptable results, while our 2BiVQA model achieves the second-highest performance, outperforming the majority of recent \ac{sota} models, with FAST-VQA ranking as the best performer. On LIVE-VQC, which contains many mobile videos showing huge camera motions, 2BiVQA consistently ranks within the top three performers based on evaluation metrics. \ac{tlvqm} method also yields competitive scores on this dataset, thanks to its many heavily designed motion-relevant features. On YouTube-UGC, RAPIQUE, 2BiVQA, FAST-VQA and \ac{videval} metrics achieve the best correlation scores, outperforming by fare the other \ac{bvqa} models. Finally, for the largest dataset (All-Combined), 2BiVQA delivers the third-highest performance, slightly outperformed by \ac{rapique} and FAST-VQA.

Although our 2BiVQA method does not outperform FAST-VQA in terms of correlation metrics, it has significant advantages in terms of training efficiency. Table \ref{com} provides a comparison between the characteristics of 2BiVQA and FAST-VQA  during training. Notably, 2BiVQA shows approximately 15 times faster training time than FAST-VQA.  Furthermore, our approach is efficient in terms of energy consumption, using approximately 30 times less energy than FAST-VQA. In addition, 2BiVQA has fewer model parameters during training than FAST-VQA, thereby simplifying the training process and improving the efficiency of resource allocation.

\begin{table}[t!]
\small
\centering
\caption{Cross dataset generalization in terms of SROCC.}
\adjustbox{max width=0.5\textwidth}{%
\label{table5}
\begin{tabular}{@{}
>{\columncolor[HTML]{FFFFFF}}l 
>{\columncolor[HTML]{FFFFFF}}l 
>{\columncolor[HTML]{FFFFFF}}c 
>{\columncolor[HTML]{FFFFFF}}c 
>{\columncolor[HTML]{FFFFFF}}c @{}}
\toprule
Model                                            & Train \textbackslash Test & KoNViD-1K       & LIVE-VQC & YouTube-UGC \\ \midrule
\cellcolor[HTML]{FFFFFF}                          & KoNViD-1K    & -          & 0.604   & 0.392      \\
\cellcolor[HTML]{FFFFFF}                          & LIVE-VQC     & 0.644          & -   & 0.277      \\
\multirow{-3}{*}{\cellcolor[HTML]{FFFFFF}VIDEVAL}                          & YouTube-UGC  & 0.594          & 0.388   &  -      \\
 \midrule
\cellcolor[HTML]{FFFFFF}                          & KoNViD-1K    & -          & 0.546   & 0.318      \\
\cellcolor[HTML]{FFFFFF}                          & LIVE-VQC     & 0.656          & -   & 0.352      \\
\multirow{-3}{*}{\cellcolor[HTML]{FFFFFF}RAPIQUE}                          & YouTube-UGC  & 0.582          & 0.623   & -      \\
 \midrule
\cellcolor[HTML]{FFFFFF}                          & KoNViD-1K    & -          & 0.734   & 0.373      \\
\cellcolor[HTML]{FFFFFF}                          & LIVE-VQC     & 0.750          & -   & 0.365      \\
\multirow{-3}{*}{\cellcolor[HTML]{FFFFFF}FAST-VQA}                          & YouTube-UGC  & \textbf{0.687}          & 0.658   & -      \\
 \midrule
\cellcolor[HTML]{FFFFFF}                          & KoNViD-1K    &  -          & \textbf{0.770}   & \textbf{0.428}      \\
\cellcolor[HTML]{FFFFFF}                          & LIVE-VQC     & \textbf{0.753} & -   & \textbf{0.416}      \\
\multirow{-3}{*}{\cellcolor[HTML]{FFFFFF}2BiVQA}                          & YouTube-UGC  & 0.647       & \textbf{0.674}   & -      \\ \bottomrule 
\end{tabular}
}
\end{table}

Figure \ref{scatter} shows  the \ac{mos} versus the prediction scores and nonlinear logistic fitted curves for the three best performing models (\ac{videval}, RAPIQUE and 2BiVQA) on the four evaluated datasets. These figures illustrate visually that the performance of 2BiVQA remains stable over the different datasets. Its scatter points are more densely clustered around the fitted curves, which are also more linear, especially for KoNViD-1k, YouTube-UGC, and All-Combined datasets.

\subsection{Cross dataset generalization}
A good \ac{vqa} metric is supposed to generalize to unseen samples. Accordingly, we perform a cross-dataset evaluation by training the three best performing \ac{bvqa} models on one dataset and testing them on the other datasets. The results are shown in Table~\ref{table5}. From this table, we can observe that the proposed model generalizes well to unseen datasets, and its performance does not depend on the dataset, which represents an essential feature for \ac{ugc}-\ac{bvqa}. Notably, 2BiVQA demonstrates superior generalization capability compared to FAST-VQA, RAPIQUE, and VIDEVAL, as indicated by the obtained results. This good generalization of the proposed method, which we believe is primarily due to the separation of the training into two stages, first the pre-training on KonIQ-10k dataset and then the fine-tuning on the target \ac{ugc}-\ac{vqa} dataset. The training on this diverse content allows our model to learn a rich feature representation suitable for \ac{ugc} video quality score prediction. It can also be noted that the cross-domain \ac{bvqa} methods generalization using YouTube-UGC is the best on average.

\begin{table}[t!]
\centering
\caption{Average runtime comparison evaluated on 1080p videos from YouTube-UGC.}
\label{table6}
\adjustbox{max width=0.5\textwidth}{%
\begin{tabular}{@{}
>{\columncolor[HTML]{FFFFFF}}l 
>{\columncolor[HTML]{FFFFFF}}c 
>{\columncolor[HTML]{FFFFFF}}c 
>{\columncolor[HTML]{FFFFFF}}c 
>{\columncolor[HTML]{FFFFFF}}c @{}}
\toprule 
\cellcolor[HTML]{FFFFFF}                          & \multicolumn{1}{l}{\cellcolor[HTML]{FFFFFF}}                                & \multicolumn{1}{l}{\cellcolor[HTML]{FFFFFF}}                            & \multicolumn{2}{l}{\cellcolor[HTML]{FFFFFF}Time (Sec.)} \\ \cmidrule(l){4-5} 
\multirow{-2}{*}{\cellcolor[HTML]{FFFFFF}Method} & \multicolumn{1}{l}{\multirow{-2}{*}{\cellcolor[HTML]{FFFFFF}Deep Learning}} & \multicolumn{1}{l}{\multirow{-2}{*}{\cellcolor[HTML]{FFFFFF}Framework}} & CPU                        & GPU                       \\ \midrule
BRISQUE                                           & \xmark                                                                       & MATLAB                                                                  & 1.45                       & \xmark                     \\
NIQE                                              & \xmark                                                                       & MATLAB                                                                  & 1.68                       & \xmark                     \\
GM-LOG                                            & \xmark                                                                       & MATLAB                                                                  & 1.77                       & \xmark                     \\
VIDEVAL                                           & \xmark                                                                       & MATLAB                                                                  & 217.2                      & \xmark                     \\
RAPIQUE                                           & \cmark                                                                       & MATLAB                                                                  & 12.6                       & \xmark                     \\
VGG19                                             & \cmark                                                                       & TensorFlow                                                              & 9.26                       & 7.81                      \\
FAST-VQA                                            & \cmark                                                                       & PyTorch                                                              & 13.4                      & 4.9                      \\ 
2BiVQA                                            & \cmark                                                                       & TensorFlow                                                              & 16.2                      & 13.6                      \\
\bottomrule
\end{tabular}
}
\end{table}

\begin{figure}[!]
  \centering
    \includegraphics[width=\linewidth]{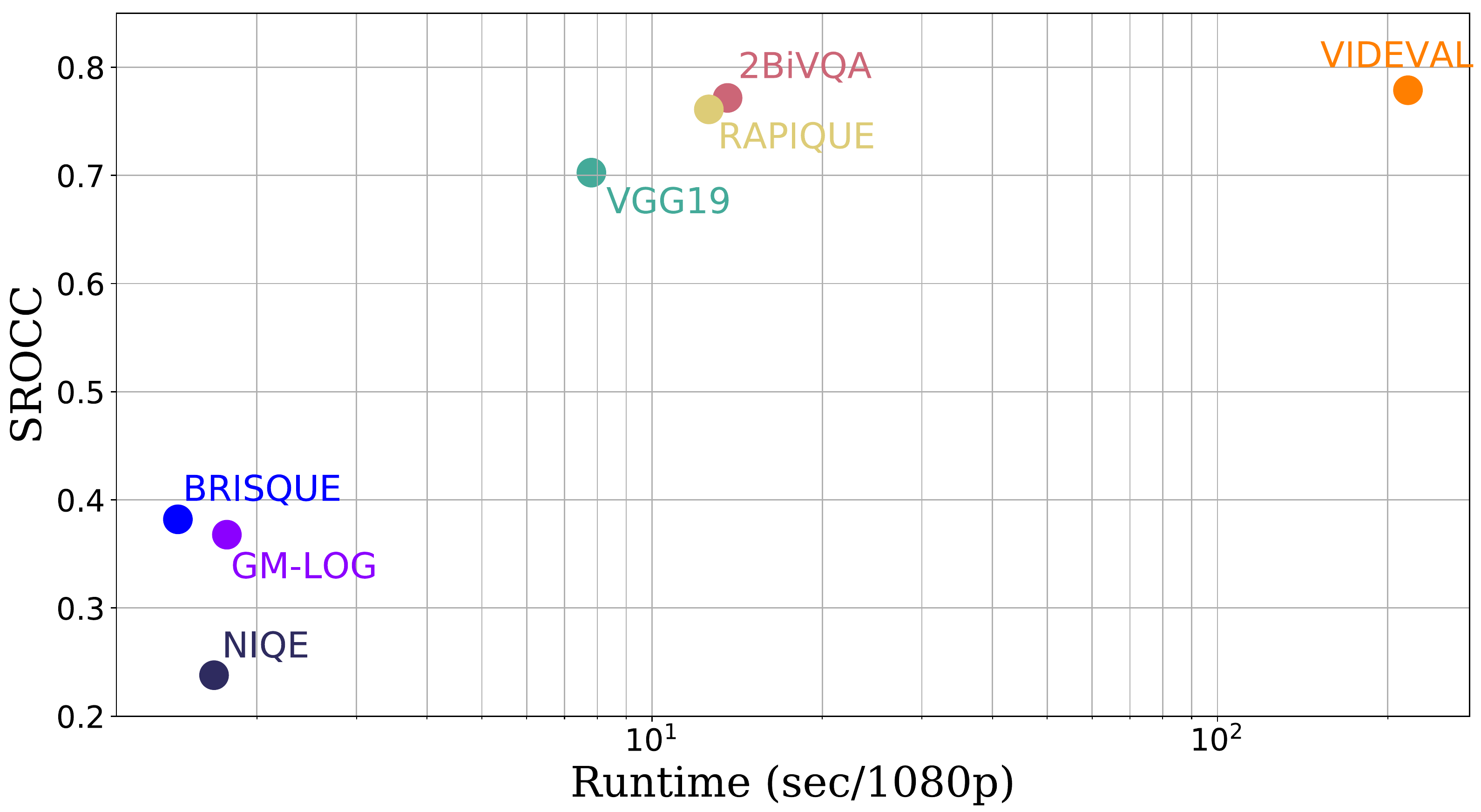}
\caption{Scatter plots of SRCC (on YouTube-UGC) of selected \ac{bvqa} methods versus runtime (on 1080p).}
\label{runtime_vs_srocc}
\end{figure}

\subsection{Complexity and runtime comparison}

Computational efficiency is crucial for \ac{vqa} algorithms, especially in practical deployments. In this regard, we performed runtime comparisons of our model as well as several methods on the same desktop computer equipped with an Intel® Xeon W-2145 CPU @ 3.70GHz $\times$16, 64G RAM, and GeForce RTX 2080 Ti graphics card under Ubuntu 20.04 \ac{lts} operating system. We used the initially released implementation in MATLAB R2018b and python 3.8.8 for GM-LOG, \ac{videval}, and RAPIQUE metrics. For BRISQUE and NIQE, we used scikit-video python library implementation. FAST-VQA was implemented in PyTorch and the remaining models, namely VGG19 and 2BiVQA, were implemented in TensorFlow. All \ac{biqa} models extract features at one frame per second, and then an average pooling was used to get the overall video quality score. We consider videos from YouTube-UGC at HD resolution ($1920 \times 1080$), then we recorded the average runtime in seconds, as shown in Table \ref{table6}. For better illustration, Figure \ref{runtime_vs_srocc} shows the scatter plots of \ac{srocc} versus runtime. It may be observed that FAST-VQA and \ac{biqa} models are faster than other methods, while VGG19, RAPIQUE, and 2BiVQA are relatively comparable.

\section{Conclusion}
\label{cncls}
In this paper, we proposed an effective \ac{bvqa} metric for \ac{ugc} videos, named 2BiVQA for double Bi-LSTM Video Quality Assessment. Our contribution relies on a deep \ac{cnn}-based model to extract frame-level features and two \ac{bi-lstm} networks for spatial and temporal pooling. Specifically, the first \ac{bi-lstm} network is used to eﬀiciently capture the short-term dependencies between neighboring patches, while the second \ac{bi-lstm} network is exploited to capture long-range dependencies between frames over the entire video. In this way, the proposed 2BiVQA can take into account the features of \ac{ugc} videos and mimic the behavior of the  \ac{hvs}. 

In addition, the training was carried out in two stages to avoid over-fitting with the limited dataset. This training strategy improved feature representation, which significantly increased accuracy performance.

We conducted comprehensive tests on four \ac{ugc}-\ac{vqa} datasets. Results showed that 2BiVQA outperforms \ac{sota} methods on two of the considered datasets (KonViD-1k and LIVE-VQC) and achieves competitive performance on YouTube-UGC and All-Combined. We further showed that the performance of the proposed solution is independent of the training dataset and generalizes better on unseen datasets than other \ac{bvqa} methods, which is a key feature of the \ac{ugc} \ac{vqa} problem. Finally, since computational efficiency is crucial for \ac{bvqa} algorithms, 2BiVQA has achieved a good trade-off between inference runtime, prediction performance and model complexity.

One future work worth addressing is to extend 2BiVQA to UGC AVQA. This can be achieved by incorporating the audio information into the spatial and temporal pooling blocs.

\ifCLASSOPTIONcaptionsoff
  \newpage
\fi



%




\bibliographystyle{IEEEtran}
\bibliography{IEEEabrv}

\begin{thebibliography}{100}
\providecommand{\url}[1]{#1}
\csname url@samestyle\endcsname
\providecommand{\newblock}{\relax}
\providecommand{\bibinfo}[2]{#2}
\providecommand{\BIBentrySTDinterwordspacing}{\spaceskip=0pt\relax}
\providecommand{\BIBentryALTinterwordstretchfactor}{4}
\providecommand{\BIBentryALTinterwordspacing}{\spaceskip=\fontdimen2\font plus
\BIBentryALTinterwordstretchfactor\fontdimen3\font minus
  \fontdimen4\font\relax}
\providecommand{\BIBforeignlanguage}[2]{{%
\expandafter\ifx\csname l@#1\endcsname\relax
\typeout{** WARNING: IEEEtran.bst: No hyphenation pattern has been}%
\typeout{** loaded for the language `#1'. Using the pattern for}%
\typeout{** the default language instead.}%
\else
\language=\csname l@#1\endcsname
\fi
#2}}
\providecommand{\BIBdecl}{\relax}
\BIBdecl

\bibitem{cisco2020cisco}
U.~Cisco, ``Cisco annual internet report (2018--2023) white paper,''
  \emph{Cisco: San Jose, CA, USA}, 2020.

\bibitem{bt2012500}
I.~R. BT, ``500-13,“,'' \emph{Methodology for the subjective assessment of
  the quality of television pictures,” International Telecommunication
  Union}, vol.~6, 2012.

\bibitem{seshadrinathan2010study}
K.~Seshadrinathan, R.~Soundararajan, A.~C. Bovik, and L.~K. Cormack, ``Study of
  subjective and objective quality assessment of video,'' \emph{IEEE
  transactions on Image Processing}, vol.~19, no.~6, pp. 1427--1441, 2010.

\bibitem{tu2021ugc}
Z.~Tu, Y.~Wang, N.~Birkbeck, B.~Adsumilli, and A.~C. Bovik, ``Ugc-vqa:
  Benchmarking blind video quality assessment for user generated content,''
  \emph{IEEE Transactions on Image Processing}, vol.~30, pp. 4449--4464, 2021.

\bibitem{hosu2017konstanz}
V.~Hosu, F.~Hahn, M.~Jenadeleh, H.~Lin, H.~Men, T.~Szir{\'a}nyi, S.~Li, and
  D.~Saupe, ``The konstanz natural video database (konvid-1k),'' in \emph{2017
  Ninth international conference on quality of multimedia experience
  (QoMEX)}.\hskip 1em plus 0.5em minus 0.4em\relax IEEE, 2017, pp. 1--6.

\bibitem{sinno2018large}
Z.~Sinno and A.~C. Bovik, ``Large-scale study of perceptual video quality,''
  \emph{IEEE Transactions on Image Processing}, vol.~28, no.~2, pp. 612--627,
  2018.

\bibitem{wang2019youtube}
Y.~Wang, S.~Inguva, and B.~Adsumilli, ``Youtube ugc dataset for video
  compression research,'' in \emph{2019 IEEE 21st International Workshop on
  Multimedia Signal Processing (MMSP)}.\hskip 1em plus 0.5em minus 0.4em\relax
  IEEE, 2019, pp. 1--5.

\bibitem{xu2021perceptual}
J.~Xu, J.~Li, X.~Zhou, W.~Zhou, B.~Wang, and Z.~Chen, ``Perceptual quality
  assessment of internet videos,'' in \emph{Proceedings of the 29th ACM
  International Conference on Multimedia}, 2021, pp. 1248--1257.

\bibitem{zhang2015aesthetics}
Y.~Zhang, L.~Zhang, and R.~Zimmermann, ``Aesthetics-guided summarization from
  multiple user generated videos,'' \emph{ACM Transactions on Multimedia
  Computing, Communications, and Applications (TOMM)}, vol.~11, no.~2, pp.
  1--23, 2015.

\bibitem{marziliano2002no}
P.~Marziliano, F.~Dufaux, S.~Winkler, and T.~Ebrahimi, ``A no-reference
  perceptual blur metric,'' in \emph{Proceedings. International conference on
  image processing}, vol.~3.\hskip 1em plus 0.5em minus 0.4em\relax IEEE, 2002,
  pp. III--III.

\bibitem{wang2008blind}
X.~Wang, B.~Tian, C.~Liang, and D.~Shi, ``Blind image quality assessment for
  measuring image blur,'' in \emph{2008 Congress on Image and Signal
  Processing}, vol.~1.\hskip 1em plus 0.5em minus 0.4em\relax IEEE, 2008, pp.
  467--470.

\bibitem{wang2000blind}
Z.~Wang, A.~C. Bovik, and B.~L. Evan, ``Blind measurement of blocking artifacts
  in images,'' in \emph{Proceedings 2000 International Conference on Image
  Processing (Cat. No. 00CH37101)}, vol.~3.\hskip 1em plus 0.5em minus
  0.4em\relax Ieee, 2000, pp. 981--984.

\bibitem{min2017unified}
X.~Min, K.~Ma, K.~Gu, G.~Zhai, Z.~Wang, and W.~Lin, ``Unified blind quality
  assessment of compressed natural, graphic, and screen content images,''
  \emph{IEEE Transactions on Image Processing}, vol.~26, no.~11, pp.
  5462--5474, 2017.

\bibitem{feng2006measurement}
X.~Feng and J.~P. Allebach, ``Measurement of ringing artifacts in jpeg
  images,'' in \emph{Digital Publishing}, vol. 6076.\hskip 1em plus 0.5em minus
  0.4em\relax International Society for Optics and Photonics, 2006, p. 60760A.

\bibitem{wang2016perceptual}
Y.~Wang, S.-U. Kum, C.~Chen, and A.~Kokaram, ``A perceptual visibility metric
  for banding artifacts,'' in \emph{2016 IEEE International Conference on Image
  Processing (ICIP)}.\hskip 1em plus 0.5em minus 0.4em\relax IEEE, 2016, pp.
  2067--2071.

\bibitem{tu2020bband}
Z.~Tu, J.~Lin, Y.~Wang, B.~Adsumilli, and A.~C. Bovik, ``Bband index: a
  no-reference banding artifact predictor,'' in \emph{ICASSP 2020-2020 IEEE
  International Conference on Acoustics, Speech and Signal Processing
  (ICASSP)}.\hskip 1em plus 0.5em minus 0.4em\relax IEEE, 2020, pp. 2712--2716.

\bibitem{amer2005fast}
A.~Amer and E.~Dubois, ``Fast and reliable structure-oriented video noise
  estimation,'' \emph{IEEE Transactions on Circuits and Systems for Video
  Technology}, vol.~15, no.~1, pp. 113--118, 2005.

\bibitem{norkin2018film}
A.~Norkin and N.~Birkbeck, ``Film grain synthesis for av1 video codec,'' in
  \emph{2018 Data Compression Conference}.\hskip 1em plus 0.5em minus
  0.4em\relax IEEE, 2018, pp. 3--12.

\bibitem{gu2014hybrid}
K.~Gu, G.~Zhai, X.~Yang, and W.~Zhang, ``Hybrid no-reference quality metric for
  singly and multiply distorted images,'' \emph{IEEE Transactions on
  Broadcasting}, vol.~60, no.~3, pp. 555--567, 2014.

\bibitem{lu2015no}
Y.~Lu, F.~Xie, T.~Liu, Z.~Jiang, and D.~Tao, ``No reference quality assessment
  for multiply-distorted images based on an improved bag-of-words model,''
  \emph{IEEE Signal Processing Letters}, vol.~22, no.~10, pp. 1811--1815, 2015.

\bibitem{min2018blind}
X.~Min, G.~Zhai, K.~Gu, Y.~Liu, and X.~Yang, ``Blind image quality estimation
  via distortion aggravation,'' \emph{IEEE Transactions on Broadcasting},
  vol.~64, no.~2, pp. 508--517, 2018.

\bibitem{zhai2020perceptual}
G.~Zhai and X.~Min, ``Perceptual image quality assessment: a survey,''
  \emph{Science China Information Sciences}, vol.~63, pp. 1--52, 2020.

\bibitem{min2021screen}
X.~Min, K.~Gu, G.~Zhai, X.~Yang, W.~Zhang, P.~Le~Callet, and C.~W. Chen,
  ``Screen content quality assessment: overview, benchmark, and beyond,''
  \emph{ACM Computing Surveys (CSUR)}, vol.~54, no.~9, pp. 1--36, 2021.

\bibitem{moorthy2011blind}
A.~K. Moorthy and A.~C. Bovik, ``Blind image quality assessment: From natural
  scene statistics to perceptual quality,'' \emph{IEEE transactions on Image
  Processing}, vol.~20, no.~12, pp. 3350--3364, 2011.

\bibitem{kundu2017no}
D.~Kundu, D.~Ghadiyaram, A.~C. Bovik, and B.~L. Evans, ``No-reference quality
  assessment of tone-mapped hdr pictures,'' \emph{IEEE Transactions on Image
  Processing}, vol.~26, no.~6, pp. 2957--2971, 2017.

\bibitem{ghadiyaram2017perceptual}
D.~Ghadiyaram and A.~C. Bovik, ``Perceptual quality prediction on authentically
  distorted images using a bag of features approach,'' \emph{Journal of
  vision}, vol.~17, no.~1, pp. 32--32, 2017.

\bibitem{pei2015image}
S.-C. Pei and L.-H. Chen, ``Image quality assessment using human visual dog
  model fused with random forest,'' \emph{IEEE Transactions on Image
  Processing}, vol.~24, no.~11, pp. 3282--3292, 2015.

\bibitem{ruderman1994statistics}
D.~L. Ruderman and W.~Bialek, ``Statistics of natural images: Scaling in the
  woods,'' \emph{Physical review letters}, vol.~73, no.~6, p. 814, 1994.

\bibitem{mittal2012no}
A.~Mittal, A.~K. Moorthy, and A.~C. Bovik, ``No-reference image quality
  assessment in the spatial domain,'' \emph{IEEE Transactions on image
  processing}, vol.~21, no.~12, pp. 4695--4708, 2012.

\bibitem{saad2010dct}
M.~A. Saad, A.~C. Bovik, and C.~Charrier, ``A dct statistics-based blind image
  quality index,'' \emph{IEEE Signal Processing Letters}, vol.~17, no.~6, pp.
  583--586, 2010.

\bibitem{saad2012blind}
------, ``Blind image quality assessment: A natural scene statistics approach
  in the dct domain,'' \emph{IEEE transactions on Image Processing}, vol.~21,
  no.~8, pp. 3339--3352, 2012.

\bibitem{moorthy2010two}
A.~K. Moorthy and A.~C. Bovik, ``A two-step framework for constructing blind
  image quality indices,'' \emph{IEEE Signal processing letters}, vol.~17,
  no.~5, pp. 513--516, 2010.

\bibitem{zhang2014c}
Y.~Zhang, A.~K. Moorthy, D.~M. Chandler, and A.~C. Bovik, ``C-diivine:
  No-reference image quality assessment based on local magnitude and phase
  statistics of natural scenes,'' \emph{Signal processing: image
  communication}, vol.~29, no.~7, pp. 725--747, 2014.

\bibitem{li2016spatiotemporal}
X.~Li, Q.~Guo, and X.~Lu, ``Spatiotemporal statistics for video quality
  assessment,'' \emph{IEEE Transactions on Image Processing}, vol.~25, no.~7,
  pp. 3329--3342, 2016.

\bibitem{sinno2019spatio}
Z.~Sinno and A.~C. Bovik, ``Spatio-temporal measures of naturalness,'' in
  \emph{2019 IEEE International Conference on Image Processing (ICIP)}.\hskip
  1em plus 0.5em minus 0.4em\relax IEEE, 2019, pp. 1750--1754.

\bibitem{saad2014blind}
M.~A. Saad, A.~C. Bovik, and C.~Charrier, ``Blind prediction of natural video
  quality,'' \emph{IEEE Transactions on Image Processing}, vol.~23, no.~3, pp.
  1352--1365, 2014.

\bibitem{liu2020blind}
Y.~Liu, K.~Gu, X.~Li, and Y.~Zhang, ``Blind image quality assessment by natural
  scene statistics and perceptual characteristics,'' \emph{ACM Transactions on
  Multimedia Computing, Communications, and Applications (TOMM)}, vol.~16,
  no.~3, pp. 1--91, 2020.

\bibitem{zhang2013no}
Y.~Zhang and D.~M. Chandler, ``No-reference image quality assessment based on
  log-derivative statistics of natural scenes,'' \emph{Journal of Electronic
  Imaging}, vol.~22, no.~4, p. 043025, 2013.

\bibitem{xue2014blind}
W.~Xue, X.~Mou, L.~Zhang, A.~C. Bovik, and X.~Feng, ``Blind image quality
  assessment using joint statistics of gradient magnitude and laplacian
  features,'' \emph{IEEE Transactions on Image Processing}, vol.~23, no.~11,
  pp. 4850--4862, 2014.

\bibitem{ye2012unsupervised}
P.~Ye, J.~Kumar, L.~Kang, and D.~Doermann, ``Unsupervised feature learning
  framework for no-reference image quality assessment,'' in \emph{2012 IEEE
  conference on computer vision and pattern recognition}.\hskip 1em plus 0.5em
  minus 0.4em\relax IEEE, 2012, pp. 1098--1105.

\bibitem{xu2014no}
J.~Xu, P.~Ye, Y.~Liu, and D.~Doermann, ``No-reference video quality assessment
  via feature learning,'' in \emph{2014 IEEE international conference on image
  processing (ICIP)}.\hskip 1em plus 0.5em minus 0.4em\relax IEEE, 2014, pp.
  491--495.

\bibitem{min2017blind}
X.~Min, K.~Gu, G.~Zhai, J.~Liu, X.~Yang, and C.~W. Chen, ``Blind quality
  assessment based on pseudo-reference image,'' \emph{IEEE Transactions on
  Multimedia}, vol.~20, no.~8, pp. 2049--2062, 2017.

\bibitem{seshadrinathan2011temporal}
K.~Seshadrinathan and A.~C. Bovik, ``Temporal hysteresis model of time varying
  subjective video quality,'' in \emph{2011 IEEE international conference on
  acoustics, speech and signal processing (ICASSP)}.\hskip 1em plus 0.5em minus
  0.4em\relax IEEE, 2011, pp. 1153--1156.

\bibitem{tu2020comparative}
Z.~Tu, C.-J. Chen, L.-H. Chen, N.~Birkbeck, B.~Adsumilli, and A.~C. Bovik, ``A
  comparative evaluation of temporal pooling methods for blind video quality
  assessment,'' in \emph{2020 IEEE International Conference on Image Processing
  (ICIP)}.\hskip 1em plus 0.5em minus 0.4em\relax IEEE, 2020, pp. 141--145.

\bibitem{mittal2015completely}
A.~Mittal, M.~A. Saad, and A.~C. Bovik, ``A completely blind video integrity
  oracle,'' \emph{IEEE Transactions on Image Processing}, vol.~25, no.~1, pp.
  289--300, 2015.

\bibitem{manasa2016optical}
K.~Manasa and S.~S. Channappayya, ``An optical flow-based no-reference video
  quality assessment algorithm,'' in \emph{2016 IEEE International Conference
  on Image Processing (ICIP)}.\hskip 1em plus 0.5em minus 0.4em\relax IEEE,
  2016, pp. 2400--2404.

\bibitem{korhonen2019two}
J.~Korhonen, ``Two-level approach for no-reference consumer video quality
  assessment,'' \emph{IEEE Transactions on Image Processing}, vol.~28, no.~12,
  pp. 5923--5938, 2019.

\bibitem{9633248}
P.~Kancharla and S.~S. Channappayya, ``Completely blind quality assessment of
  user generated video content,'' \emph{IEEE Transactions on Image Processing},
  vol.~31, pp. 263--274, 2022.

\bibitem{tiotsop2022mimicking}
L.~F. Tiotsop, T.~Mizdos, M.~Barkowsky, P.~Pocta, A.~Servetti, and E.~Masala,
  ``Mimicking individual media quality perception with neural network based
  artificial observers,'' \emph{ACM Transactions on Multimedia Computing,
  Communications, and Applications (TOMM)}, vol.~18, no.~1, pp. 1--25, 2022.

\bibitem{zhu2018measuring}
Y.~Zhu, S.~C. Guntuku, W.~Lin, G.~Ghinea, and J.~A. Redi, ``Measuring
  individual video qoe: A survey, and proposal for future directions using
  social media,'' \emph{ACM Transactions on Multimedia Computing,
  Communications, and Applications (TOMM)}, vol.~14, no.~2s, pp. 1--24, 2018.

\bibitem{simonyan2014very}
K.~Simonyan and A.~Zisserman, ``Very deep convolutional networks for
  large-scale image recognition,'' in \emph{3rd International Conference on
  Learning Representations, {ICLR} 2015, San Diego, CA, USA, May 7-9, 2015,
  Conference Track Proceedings}, Y.~Bengio and Y.~LeCun, Eds., 2015.

\bibitem{he2016deep}
K.~He, X.~Zhang, S.~Ren, and J.~Sun, ``Deep residual learning for image
  recognition,'' in \emph{Proceedings of the IEEE conference on computer vision
  and pattern recognition}, 2016, pp. 770--778.

\bibitem{bochkovskiy2020yolov4}
\BIBentryALTinterwordspacing
A.~Bochkovskiy, C.~Wang, and H.~M. Liao, ``Yolov4: Optimal speed and accuracy
  of object detection,'' \emph{CoRR}, vol. abs/2004.10934, 2020. [Online].
  Available: \url{https://arxiv.org/abs/2004.10934}
\BIBentrySTDinterwordspacing

\bibitem{girshick2015fast}
R.~Girshick, ``Fast r-cnn,'' in \emph{Proceedings of the IEEE international
  conference on computer vision}, 2015, pp. 1440--1448.

\bibitem{ronneberger2015u}
O.~Ronneberger, P.~Fischer, and T.~Brox, ``U-net: Convolutional networks for
  biomedical image segmentation,'' in \emph{International Conference on Medical
  image computing and computer-assisted intervention}.\hskip 1em plus 0.5em
  minus 0.4em\relax Springer, 2015, pp. 234--241.

\bibitem{liu2015parsenet}
W.~Liu, A.~Rabinovich, and A.~C. Berg, ``Parsenet: Looking wider to see
  better,'' \emph{CoRR}, vol. abs/1506.04579, 2015.

\bibitem{ghadiyaram2015massive}
D.~Ghadiyaram and A.~C. Bovik, ``Massive online crowdsourced study of
  subjective and objective picture quality,'' \emph{IEEE Transactions on Image
  Processing}, vol.~25, no.~1, pp. 372--387, 2015.

\bibitem{gotz2021konvid}
F.~G{\"o}tz-Hahn, V.~Hosu, H.~Lin, and D.~Saupe, ``Konvid-150k: A dataset for
  no-reference video quality assessment of videos in-the-wild,'' \emph{IEEE
  Access}, vol.~9, pp. 72\,139--72\,160, 2021.

\bibitem{kang2014convolutional}
L.~Kang, P.~Ye, Y.~Li, and D.~Doermann, ``Convolutional neural networks for
  no-reference image quality assessment,'' in \emph{Proceedings of the IEEE
  conference on computer vision and pattern recognition}, 2014, pp. 1733--1740.

\bibitem{kim2017deep}
J.~Kim, H.~Zeng, D.~Ghadiyaram, S.~Lee, L.~Zhang, and A.~C. Bovik, ``Deep
  convolutional neural models for picture-quality prediction: Challenges and
  solutions to data-driven image quality assessment,'' \emph{IEEE Signal
  processing magazine}, vol.~34, no.~6, pp. 130--141, 2017.

\bibitem{deng2009imagenet}
J.~Deng, W.~Dong, R.~Socher, L.-J. Li, K.~Li, and L.~Fei-Fei, ``Imagenet: A
  large-scale hierarchical image database,'' in \emph{2009 IEEE conference on
  computer vision and pattern recognition}.\hskip 1em plus 0.5em minus
  0.4em\relax Ieee, 2009, pp. 248--255.

\bibitem{yang2019survey}
X.~Yang, F.~Li, and H.~Liu, ``A survey of dnn methods for blind image quality
  assessment,'' \emph{IEEE Access}, vol.~7, pp. 123\,788--123\,806, 2019.

\bibitem{sun2023blind}
W.~Sun, X.~Min, D.~Tu, S.~Ma, and G.~Zhai, ``Blind quality assessment for
  in-the-wild images via hierarchical feature fusion and iterative mixed
  database training,'' \emph{IEEE Journal of Selected Topics in Signal
  Processing}, 2023.

\bibitem{ahn2018deep}
S.~Ahn and S.~Lee, ``Deep blind video quality assessment based on temporal
  human perception,'' in \emph{2018 25th IEEE International Conference on Image
  Processing (ICIP)}.\hskip 1em plus 0.5em minus 0.4em\relax IEEE, 2018, pp.
  619--623.

\bibitem{you2019deep}
J.~You and J.~Korhonen, ``Deep neural networks for no-reference video quality
  assessment,'' in \emph{2019 IEEE International Conference on Image Processing
  (ICIP)}.\hskip 1em plus 0.5em minus 0.4em\relax IEEE, 2019, pp. 2349--2353.

\bibitem{liu2018end}
W.~Liu, Z.~Duanmu, and Z.~Wang, ``End-to-end blind quality assessment of
  compressed videos using deep neural networks.'' in \emph{ACM Multimedia},
  2018, pp. 546--554.

\bibitem{li2019quality}
D.~Li, T.~Jiang, and M.~Jiang, ``Quality assessment of in-the-wild videos,'' in
  \emph{Proceedings of the 27th ACM International Conference on Multimedia},
  2019, pp. 2351--2359.

\bibitem{li2021unified}
------, ``Unified quality assessment of in-the-wild videos with mixed datasets
  training,'' \emph{International Journal of Computer Vision}, vol. 129, no.~4,
  pp. 1238--1257, 2021.

\bibitem{yi2021attention}
F.~Yi, M.~Chen, W.~Sun, X.~Min, Y.~Tian, and G.~Zhai, ``Attention based network
  for no-reference ugc video quality assessment,'' in \emph{2021 IEEE
  International Conference on Image Processing (ICIP)}.\hskip 1em plus 0.5em
  minus 0.4em\relax IEEE, 2021, pp. 1414--1418.

\bibitem{zhang2023md}
Z.~Zhang, W.~Wu, W.~Sun, D.~Tu, W.~Lu, X.~Min, Y.~Chen, and G.~Zhai, ``Md-vqa:
  Multi-dimensional quality assessment for ugc live videos,'' in
  \emph{Proceedings of the IEEE/CVF Conference on Computer Vision and Pattern
  Recognition}, 2023, pp. 1746--1755.

\bibitem{tu2021rapique}
Z.~Tu, X.~Yu, Y.~Wang, N.~Birkbeck, B.~Adsumilli, and A.~C. Bovik, ``Rapique:
  Rapid and accurate video quality prediction of user generated content,''
  \emph{IEEE Open Journal of Signal Processing}, vol.~2, pp. 425--440, 2021.

\bibitem{sun2022deep}
W.~Sun, X.~Min, W.~Lu, and G.~Zhai, ``A deep learning based no-reference
  quality assessment model for ugc videos,'' in \emph{Proceedings of the 30th
  ACM International Conference on Multimedia}, 2022, pp. 856--865.

\bibitem{sun2021deep}
W.~Sun, T.~Wang, X.~Min, F.~Yi, and G.~Zhai, ``Deep learning based
  full-reference and no-reference quality assessment models for compressed ugc
  videos,'' in \emph{2021 IEEE International Conference on Multimedia \& Expo
  Workshops (ICMEW)}.\hskip 1em plus 0.5em minus 0.4em\relax IEEE, 2021, pp.
  1--6.

\bibitem{shen2022end}
W.~Shen, M.~Zhou, X.~Liao, W.~Jia, T.~Xiang, B.~Fang, and Z.~Shang, ``An
  end-to-end no-reference video quality assessment method with hierarchical
  spatiotemporal feature representation,'' \emph{IEEE Transactions on
  Broadcasting}, 2022.

\bibitem{mitra2022multiview}
S.~Mitra and R.~Soundararajan, ``Multiview contrastive learning for completely
  blind video quality assessment of user generated content,'' in
  \emph{Proceedings of the 30th ACM International Conference on Multimedia},
  2022, pp. 1914--1924.

\bibitem{wu2022fast}
H.~Wu, C.~Chen, J.~Hou, L.~Liao, A.~Wang, W.~Sun, Q.~Yan, and W.~Lin,
  ``Fast-vqa: Efficient end-to-end video quality assessment with fragment
  sampling,'' in \emph{European Conference on Computer Vision}.\hskip 1em plus
  0.5em minus 0.4em\relax Springer, 2022, pp. 538--554.

\bibitem{wu2023exploring}
H.~Wu, E.~Zhang, L.~Liao, C.~Chen, J.~Hou, A.~Wang, W.~Sun, Q.~Yan, and W.~Lin,
  ``Exploring video quality assessment on user generated contents from
  aesthetic and technical perspectives,'' in \emph{Proceedings of the IEEE/CVF
  International Conference on Computer Vision (ICCV)}, 2023.

\bibitem{huang2023xgc}
X.~Huang, C.~Li, A.~Bentaleb, R.~Zimmermann, and G.~Zhai, ``Xgc-vqa: A unified
  video quality assessment model for user, professionally, and
  occupationally-generated content,'' \emph{arXiv preprint arXiv:2303.13859},
  2023.

\bibitem{min2020study}
X.~Min, G.~Zhai, J.~Zhou, M.~C. Farias, and A.~C. Bovik, ``Study of subjective
  and objective quality assessment of audio-visual signals,'' \emph{IEEE
  Transactions on Image Processing}, vol.~29, pp. 6054--6068, 2020.

\bibitem{cao2023subjective}
Y.~Cao, X.~Min, W.~Sun, and G.~Zhai, ``Subjective and objective audio-visual
  quality assessment for user generated content,'' \emph{IEEE Transactions on
  Image Processing}, 2023.

\bibitem{zhang2018unreasonable}
R.~Zhang, P.~Isola, A.~A. Efros, E.~Shechtman, and O.~Wang, ``The unreasonable
  effectiveness of deep features as a perceptual metric,'' in \emph{IEEE
  conference on computer vision and pattern recognition}, 2018, pp. 586--595.

\bibitem{gao2017deepsim}
F.~Gao, Y.~Wang, P.~Li, M.~Tan, J.~Yu, and Y.~Zhu, ``Deepsim: Deep similarity
  for image quality assessment,'' \emph{Neurocomputing}, vol. 257, pp.
  104--114, 2017.

\bibitem{yang2020deep}
X.~Yang, F.~Li, and H.~Liu, ``Deep feature importance awareness based
  no-reference image quality prediction,'' \emph{Neurocomputing}, vol. 401, pp.
  209--223, 2020.

\bibitem{amirshahi2016image}
S.~A. Amirshahi, M.~Pedersen, and S.~X. Yu, ``Image quality assessment by
  comparing cnn features between images,'' \emph{Journal of Imaging Science and
  Technology}, vol.~60, no.~6, pp. 60\,410--1, 2016.

\bibitem{johnson2016perceptual}
J.~Johnson, A.~Alahi, and L.~Fei-Fei, ``Perceptual losses for real-time style
  transfer and super-resolution,'' in \emph{European conference on computer
  vision}.\hskip 1em plus 0.5em minus 0.4em\relax Springer, 2016, pp. 694--711.

\bibitem{zhang2015accelerating}
X.~Zhang, J.~Zou, K.~He, and J.~Sun, ``Accelerating very deep convolutional
  networks for classification and detection,'' \emph{IEEE transactions on
  pattern analysis and machine intelligence}, vol.~38, no.~10, pp. 1943--1955,
  2015.

\bibitem{watson1997digital}
A.~B. Watson and C.~H. Null, ``Digital images and human vision,'' in
  \emph{Electronic Imaging Science and Technology Conference}, 1997.

\bibitem{schuster1997bidirectional}
M.~Schuster and K.~K. Paliwal, ``Bidirectional recurrent neural networks,''
  \emph{IEEE transactions on Signal Processing}, vol.~45, no.~11, pp.
  2673--2681, 1997.

\bibitem{hochreiter1997long}
S.~Hochreiter and J.~Schmidhuber, ``Long short-term memory,'' \emph{Neural
  computation}, vol.~9, no.~8, pp. 1735--1780, 1997.

\bibitem{hochreiter1998vanishing}
S.~Hochreiter, ``The vanishing gradient problem during learning recurrent
  neural nets and problem solutions,'' \emph{International Journal of
  Uncertainty, Fuzziness and Knowledge-Based Systems}, vol.~6, no.~02, pp.
  107--116, 1998.

\bibitem{siami2019performance}
S.~Siami-Namini, N.~Tavakoli, and A.~S. Namin, ``The performance of lstm and
  bilstm in forecasting time series,'' in \emph{2019 IEEE International
  Conference on Big Data (Big Data)}.\hskip 1em plus 0.5em minus 0.4em\relax
  IEEE, 2019, pp. 3285--3292.

\bibitem{graves2005framewise}
A.~Graves and J.~Schmidhuber, ``Framewise phoneme classification with
  bidirectional lstm and other neural network architectures,'' \emph{Neural
  networks}, vol.~18, no. 5-6, pp. 602--610, 2005.

\bibitem{graves2013hybrid}
A.~Graves, N.~Jaitly, and A.-r. Mohamed, ``Hybrid speech recognition with deep
  bidirectional lstm,'' in \emph{2013 IEEE workshop on automatic speech
  recognition and understanding}.\hskip 1em plus 0.5em minus 0.4em\relax IEEE,
  2013, pp. 273--278.

\bibitem{hosu2020koniq}
V.~Hosu, H.~Lin, T.~Sziranyi, and D.~Saupe, ``Koniq-10k: An ecologically valid
  database for deep learning of blind image quality assessment,'' \emph{IEEE
  Transactions on Image Processing}, vol.~29, pp. 4041--4056, 2020.

\bibitem{park2012video}
J.~Park, K.~Seshadrinathan, S.~Lee, and A.~C. Bovik, ``Video quality pooling
  adaptive to perceptual distortion severity,'' \emph{IEEE Transactions on
  Image Processing}, vol.~22, no.~2, pp. 610--620, 2012.

\bibitem{kim2018deep}
W.~Kim, J.~Kim, S.~Ahn, J.~Kim, and S.~Lee, ``Deep video quality assessor: From
  spatio-temporal visual sensitivity to a convolutional neural aggregation
  network,'' in \emph{Proceedings of the European Conference on Computer Vision
  (ECCV)}, 2018, pp. 219--234.

\bibitem{ninassi2009considering}
A.~Ninassi, O.~Le~Meur, P.~Le~Callet, and D.~Barba, ``Considering temporal
  variations of spatial visual distortions in video quality assessment,''
  \emph{IEEE Journal of Selected Topics in Signal Processing}, vol.~3, no.~2,
  pp. 253--265, 2009.

\bibitem{pinson2003objective}
M.~H. Pinson and S.~Wolf, ``An objective method for combining multiple
  subjective data sets,'' in \emph{Visual Communications and Image Processing
  2003}, vol. 5150.\hskip 1em plus 0.5em minus 0.4em\relax International
  Society for Optics and Photonics, 2003, pp. 583--592.

\bibitem{DBLP:journals/corr/KingmaB14}
D.~P. Kingma and J.~Ba, ``Adam: {A} method for stochastic optimization,'' in
  \emph{3rd International Conference on Learning Representations, {ICLR} 2015,
  San Diego, CA, USA, May 7-9, 2015, Conference Track Proceedings}, Y.~Bengio
  and Y.~LeCun, Eds., 2015.

\bibitem{mittal2012making}
A.~Mittal, R.~Soundararajan, and A.~C. Bovik, ``Making a “completely blind”
  image quality analyzer,'' \emph{IEEE Signal processing letters}, vol.~20,
  no.~3, pp. 209--212, 2012.

\bibitem{zhang2015feature}
L.~Zhang, L.~Zhang, and A.~C. Bovik, ``A feature-enriched completely blind
  image quality evaluator,'' \emph{IEEE Transactions on Image Processing},
  vol.~24, no.~8, pp. 2579--2591, 2015.

\bibitem{xu2016blind}
J.~Xu, P.~Ye, Q.~Li, H.~Du, Y.~Liu, and D.~Doermann, ``Blind image quality
  assessment based on high order statistics aggregation,'' \emph{IEEE
  Transactions on Image Processing}, vol.~25, no.~9, pp. 4444--4457, 2016.

\bibitem{ying2020patches}
Z.~Ying, H.~Niu, P.~Gupta, D.~Mahajan, D.~Ghadiyaram, and A.~Bovik, ``From
  patches to pictures (paq-2-piq): Mapping the perceptual space of picture
  quality,'' in \emph{Proceedings of the IEEE/CVF Conference on Computer Vision
  and Pattern Recognition}, 2020, pp. 3575--3585.

\bibitem{huang2017densely}
G.~Huang, Z.~Liu, L.~Van Der~Maaten, and K.~Q. Weinberger, ``Densely connected
  convolutional networks,'' in \emph{Proceedings of the IEEE conference on
  computer vision and pattern recognition}, 2017, pp. 4700--4708.

\bibitem{tan2019efficientnet}
M.~Tan and Q.~Le, ``Efficientnet: Rethinking model scaling for convolutional
  neural networks,'' in \emph{International Conference on Machine
  Learning}.\hskip 1em plus 0.5em minus 0.4em\relax PMLR, 2019, pp. 6105--6114.

\end{thebibliography}

\end{document}